\documentclass[12pt, draftclsnofoot, onecolumn]{IEEEtran}

\usepackage{cite}
\usepackage[pdftex]{graphicx}
\graphicspath{{Figure/}}
\usepackage[cmex10]{amsmath}
\usepackage{amsthm}
\usepackage{amssymb}
\usepackage{algorithmic}
\usepackage{array}
\usepackage{color}
\usepackage{epstopdf}
\usepackage{amsfonts}
\usepackage{gensymb}

\usepackage{pgfplots}
\pgfplotsset{compat=1.3}
\usepackage{tikz}							
\usetikzlibrary{shapes}
\usetikzlibrary{spy}
\usetikzlibrary{circuits}
\usetikzlibrary{arrows}

\newcommand{\vect}[1]{\boldsymbol{\mathrm{#1}}}
\newcommand{\mat}[1]{\boldsymbol{\mathrm{#1}}}

\newcommand{\tr}{\mathrm{tr}}

\newcommand{\diag}{\mathrm{diag}}

\usepackage{mathtools}

\newtheorem{proposition}{Proposition}

\ifCLASSOPTIONcompsoc
  \usepackage[caption=false,font=normalsize,labelfont=sf,textfont=sf]{subfig}
\else
  \usepackage[caption=false,font=footnotesize]{subfig}
\fi
\begin{document}
%
\title{Performance Analysis of Channel Extrapolation\\ in FDD Massive MIMO Systems
}
 
\author{Fran\c{c}ois Rottenberg,~\IEEEmembership{Member,~IEEE,}
	Thomas Choi,~\IEEEmembership{Student Member,~IEEE,}\\ Peng Luo,~\IEEEmembership{Student Member,~IEEE,}
	 Jianzhong Zhang,~\IEEEmembership{Fellow,~IEEE}\\ and~Andreas F. Molisch,~\IEEEmembership{Fellow,~IEEE}
	\thanks{The work was partly supported by NSF under project ECCS-1731694 and a gift from Samsung America. The work of F. Rottenberg was also partly supported by the Belgian American Educational Foundation (B.A.E.F.). Part of the material in this paper has been presented at IEEE Globecom 2019 \cite{rottenberg2019channel}.}
	\thanks{Fran\c{c}ois Rottenberg, Thomas Choi, Peng Luo and Andreas F. Molisch are with the Ming Hsieh Department
		of Electrical and Computer Engineering, University of Southern California,
		Los Angeles, CA, USA (e-mail: \{frottenb, choit, luop, molisch\}@usc.edu). Jianzhong Zhang is with Samsung Research America, Richardson, TX, USA (e-mail: jianzhong.z@samsung.com).}
}

%



\maketitle

\begin{abstract}
Channel estimation for the downlink of frequency division duplex (FDD) massive MIMO systems is well known to generate a large overhead as the amount of training generally scales with the number of transmit antennas in a MIMO system. In this paper, we consider the solution of extrapolating the channel frequency response from uplink pilot estimates to the downlink frequency band. This drastically reduces the downlink pilot overhead and completely removes the need for a feedback from the users. The price to pay is a degradation in the quality of the channel estimates, which reduces the downlink spectral efficiency. We first show that conventional estimators fail to achieve reasonable accuracy. We propose instead to use high-resolution channel estimation. We derive the Cramer-Rao lower bound (CRLB) of the mean squared error (MSE) of the extrapolated channel. Furthermore, a relationship between the imperfect channel state information (CSI) and the downlink user performance is derived. The extrapolation-based FDD massive MIMO performance is validated through numerical simulations and compared to a corresponding time division duplex (TDD) system. Considered figures of merit for extrapolation performance include channel MSE, beamforming efficiency, extrapolation range, spectral efficiency and uncoded symbol error rate. Our main conclusion is that channel extrapolation is a viable solution for FDD massive MIMO systems.
\end{abstract}


%
\IEEEpeerreviewmaketitle


\section{Introduction}

\begin{figure*}[!t]  
	\centering
	
	\resizebox{0.95\textwidth}{!}{%
		{\includegraphics[clip, trim=0cm 8.5cm 2.5cm 0cm, scale=1]{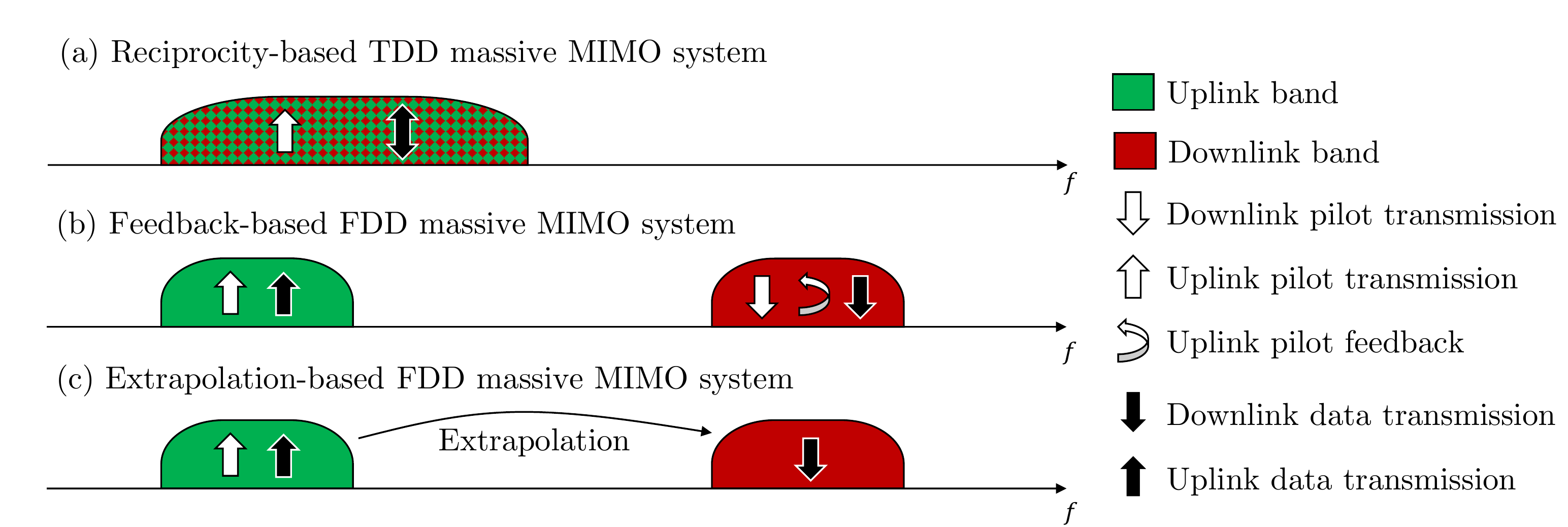}} 
	}
	\vspace{-1em}
	\caption{(a) In TDD, the BS can rely on channel reciprocity to acquire full CSI from UL pilots as UL and DL links share the same band. (b) In FDD, UL and DL bands are disjoint and reciprocity does not hold. DL pilots and their feedback is generally used to acquire DL CSI at the base station. (c) In extrapolation-based FDD, the DL CSI is directly inferred from UL pilots.}
	\label{fig:massive_mimo_TDD_FDD}
\end{figure*}

The deployment of massive multiple-input-multiple-output (MIMO) communications systems strongly relies on the acquisition of accurate channel state information (CSI) at the base station (BS) \cite{Larsson2014}. Massive MIMO systems are typically characterized by a much larger number of antennas at the BS than the sum of the antennas at the user equipments (UEs). This implies that channel estimation is much less costly in the uplink (UL) than in the downlink (DL) \cite{bjornson2017massive}. In time division duplex (TDD) systems, the BS can efficiently perform DL channel estimation from UL pilot transmission from the UEs (see Fig.~\ref{fig:massive_mimo_TDD_FDD}), since channel reciprocity holds as long as UL and DL transmission occurs within a coherence time of the channel, and {\em within the same frequency band}. However, in a frequency division duplex (FDD) scenario, reciprocity cannot be exploited as different bands, usually separated by more than a coherence bandwidth, are used in UL and DL. On the other hand, estimation of the channel by DL pilot transmission and feedback might result in a large overhead.

A variety of methods have been proposed to solve this dilemma, such as channel correlations in the spatial domain reflected in second-order statistics \cite{Adhikary2013,Barzegar2019}, compression of the feedback \cite{IEEE_80211n}, combinations thereof \cite{jiang2015achievable}, {\color{black}compression of the feedback based on deep learning \cite{Liao2019}, 3-D beamforming based on channel statistics \cite{Li2016}}, or compressed sensing methods \cite{Rao2014}, each of which involves some feedback from the users. One of the most promising methods is channel extrapolation from the UL to the DL band as it completely removes the overhead. The extrapolation range of conventional least squares (LS) and linear minimum mean squared (LMMSE) is very limited - typically to the order of one coherence bandwidth, as will be further shown in this work. To overcome this limit, \cite{Molisch2011} suggested estimation of the multipath components (MPCs) via high-resolution parameter estimation (HRPE). Based on the structure of the channel and the extracted MPCs, extrapolation over wide frequency range can be achieved. However, the paper only considered the single-input-single-output (SISO) case, which resulted in a poor extrapolation performance. Ref.~\cite{Jalden2012} extends the setup to the MIMO case and the extrapolation to the spatial domain. Multiple measurements show that a frequency extrapolation range larger than 5 times the coherence bandwidth can be reached. Ref.~\cite{Vasisht:2016} presents the so-called R2-F2 system to extract path parameters and extrapolate the channel in frequency. The paper shows how to integrate the system into LTE cellular networks and uses experimental measurements for validation. The study restricted the frequency spacing between UL and DL band to be only 20-30 MHz and did not study the mean squared error (MSE) of the extrapolated channel. Ref.~\cite{Yang2018} compares different extrapolation algorithms. This study shows that super-resolution can outperform compressed sensing methods for frequency channel extrapolation. In \cite{Zhang2018}, information about user angles is extracted from UL pilots using 2D unitary ESPRIT, a subspace-based HRPE method \cite{Haardt1995}. Then, directional training is performed in the DL. {\color{black}Ref.~\cite{Shen2018} similarly proposes an angle-of-departure (AoD) adaptive subspace codebook to reduce channel feedback overhead. In \cite{qiu2019covariance}, a hybrid statistical-instantaneous feedback mechanism where the users are separated into two classes of feedback design based on their channel covariance.} Ref.~\cite{arnold2019enabling} proposes to train a neural network to perform the channel extrapolation in frequency. This approach does not require the acquisition of the antenna array patterns through calibration but requires a large training dataset. Ref.~\cite{Ugurlu2016} proposes to acquire DL CSI through UL pilots in combination with a limited feedback from DL pilots. In \cite{choi2019}, channel extrapolation performance is experimentally evaluated, in terms of MSE of the extrapolated channel and beamforming efficiency.

Channel extrapolation in frequency also presents formal similarities to extrapolation in time. In contrast to frequency-domain extrapolation, channel prediction in time has been extensively investigated in the literature. A comprehensive review can be found in \cite{Hallen2007}. In \cite{Svantesson2006}, the authors proposed performance bounds for prediction in time of MIMO channels. They later extended their study to MIMO-OFDM channel estimation with interpolation and extrapolation being done both in time and frequency \cite{Larsen2009}. It is observed that MIMO provides much longer prediction lengths in time and frequency than for SISO systems.

To provide understanding of low-overhead FDD massive MIMO systems, this paper investigates the theoretical performance of channel extrapolation in frequency. The main originality of our paper is that it provides an in-depth theoretical study of the system performance as opposed to previous approaches, that were mostly validated through simulations and/or experiments. More specifically, we highlight the advantages of HRPE in terms of channel extrapolation as compared to conventional LS and LMMSE channel estimation. The channel MSE of both types of estimators is analytically studied. We derive the Cramer-Rao lower bound (CRLB) of the MSE, using a similar methodology as in \cite{Larsen2009}. The proposed CRLB differs from \cite{Larsen2009} by taking into account elevation angles, the frequency dependence of the pattern, and the influence of the training symbols. Furthermore, we propose a simplified CRLB, obtained under the assumption of well separated paths and giving more physical intuition about the frequency extrapolation range that can be expected in practice. The simplified CRLB shows that the MSE of the extrapolated channel frequency response is inversely proportional to the number of receive antennas while the extrapolation performance penalty scales with the square of the ratio of the frequency offset from the carrier frequency and the training bandwidth. This paper furthermore studies analytically the downlink user performance under imperfect CSI, with emphasis on the induced beamforming power loss. Finally, extensive numerical evaluations validate the extrapolation-based FDD performance and carefully compares it to a corresponding TDD system. Various performance metrics are included such as channel MSE, beamforming efficiency, extrapolation range, spectral efficiency and uncoded symbol error rate.

The rest of this paper is structured as follows. Section~\ref{section:channel_model} describes the channel model used in this work. Section~\ref{section:channel_extrapolation} introduces the LS, the LMMSE and the high-resolution SAGE estimator. Section~\ref{section:Performance_limits} studies the theoretical performance of the previously introduced estimation algorithms. Section~\ref{section:simulation_results} numerically validates the performance of an extrapolation-based FDD massive MIMO system using a standardized 3GPP channel model. Finally, Section~\ref{section:conclusion} concludes the paper and appendixes contain the mathematical proof of previous sections.

\textbf{Notations:} Vectors and matrices are denoted by bold lowercase and uppercase letters, respectively (resp.). Superscripts $^*$, $^T$ and $^H$ stand for conjugate, transpose and Hermitian transpose operators. The symbols 
$\jmath$, $\Im(.)$ and $\Re(.)$ denote the 
imaginary unit, imaginary and real parts, respectively. The expectation $\mathcal{E}[.]$ is taken over both the noise and channel statistics as opposed to $\mathbb{E}[.]$ which denotes the expectation with respect to the noise statistics only. The norm $\|.\|$ is the Frobenius norm and 
$\delta_n$ is the Kronecker delta. The $\diag(.)$ operator applied to a vector returns a diagonal matrix whose $k-$th diagonal entry is equal to the $k-$th entry of the argument vector. 

\section{Channel Model}
\label{section:channel_model}

We consider a FDD massive MIMO scenario where each user has a single-antenna and transmits an UL training sequence that is orthogonal to those of the other users. Thus, the estimations for different users become independent and in particular, the extrapolation in frequency of the SIMO channel of each user can be treated independently. For the sake of clarity and without loss of generality, we only consider one user in the following. We denote by $M$ the number of BS antennas. 

\begin{figure}[!t]  
	\centering
	
	\resizebox{0.45\textwidth}{!}{%
		{\includegraphics[clip, trim=0cm 10cm 12cm 0cm, scale=1]{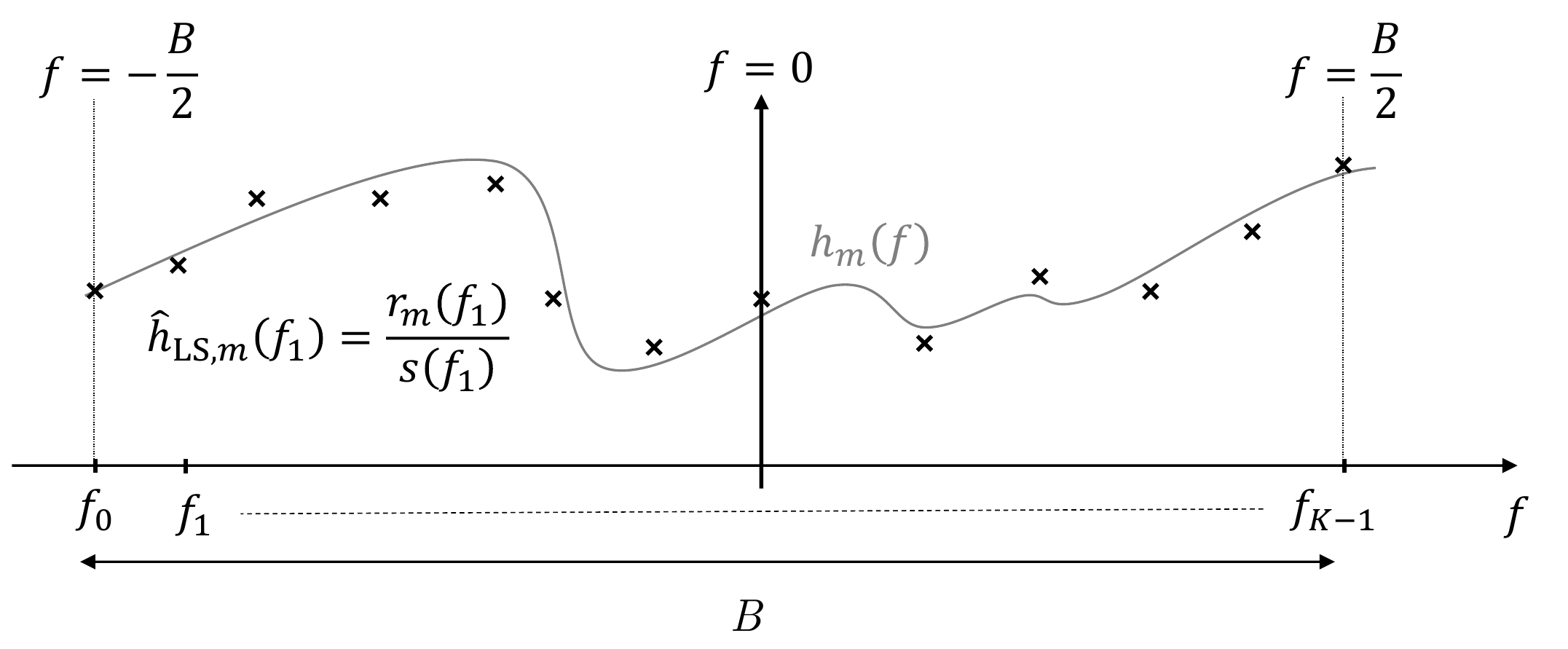}} 
	}
		\vspace{-1em}
	\caption{The base station receives $K$ pilot symbols scattered across the uplink training {\color{black} bandwidth} $B$ and at each antenna $m=1,...,M$. As a convention, the baseband frequency $f=0$ corresponds to the passband uplink carrier frequency.}
	\label{fig:system_model}
\end{figure}

We consider the transmission of a single orthogonal frequency division multiplexing (OFDM) multicarrier symbol. Since pilot symbols are orthogonal to data symbols, we only consider the samples received at pilot subcarriers for clarity. As depicted in Fig.~\ref{fig:system_model}, the BS obtains a total of $K$ pilot symbols scattered across frequency. The $k$-th transmitted pilot symbol is denoted by $s(f_k)$ with $f_k$ being the baseband frequency of the pilot subcarrier. All pilot subcarriers are transmitted inside the uplink transmission band of width $B$, \textit{i.e.}, $f_k \in [-B/2,B/2],\ k=0,...,K-1$. {\color{black}As common, we assume that the channel is time-invariant between the transmission of the uplink pilots and the use of the (extrapolated) CSI, \textit{e.g.}, for downlink beamforming. In other words, the mobility of the environment should be low enough to ensure that the coherence time of the channel is larger than this delay. Note that this quasi-static assumption is better fulfilled than in most of the FDD works in the literature involving a feedback from the user. Indeed, these works assume that the channel remains time-invariant during the transmission of downlink pilots, feedback from the users and finally the use of the obtained CSI for downlink beamforming.} The OFDM demodulated pilot symbol at antenna $m$ and frequency $f_k$ can be expressed as
\begin{align}
r_m(f_k)&=h_m(f_k)s(f_k)+w_m(f_k), \label{eq:received_signal_FD}
\end{align}
with $m=1,...,M,\ k=0,...,K-1$ and where $h_m(f_k)$ is the channel frequency response at frequency $f_k$ and antenna $m$. Samples $w_m(f_k)$ are zero mean additive complex circularly symmetric Gaussian noise of variance $\sigma_w^2$. We assume that the noise samples are uncorrelated, \textit{i.e.}, $\mathbb{E}\left(w_m(f_k)w^*_m(f_{k'})\right)=\sigma_w^2\delta_{m-m'}\delta_{k-k'}$.

\begin{figure}[!t]  
	\centering
	
	\resizebox{0.45\textwidth}{!}{%
		{\includegraphics[clip, trim=0cm 12cm 18cm 0cm, scale=1]{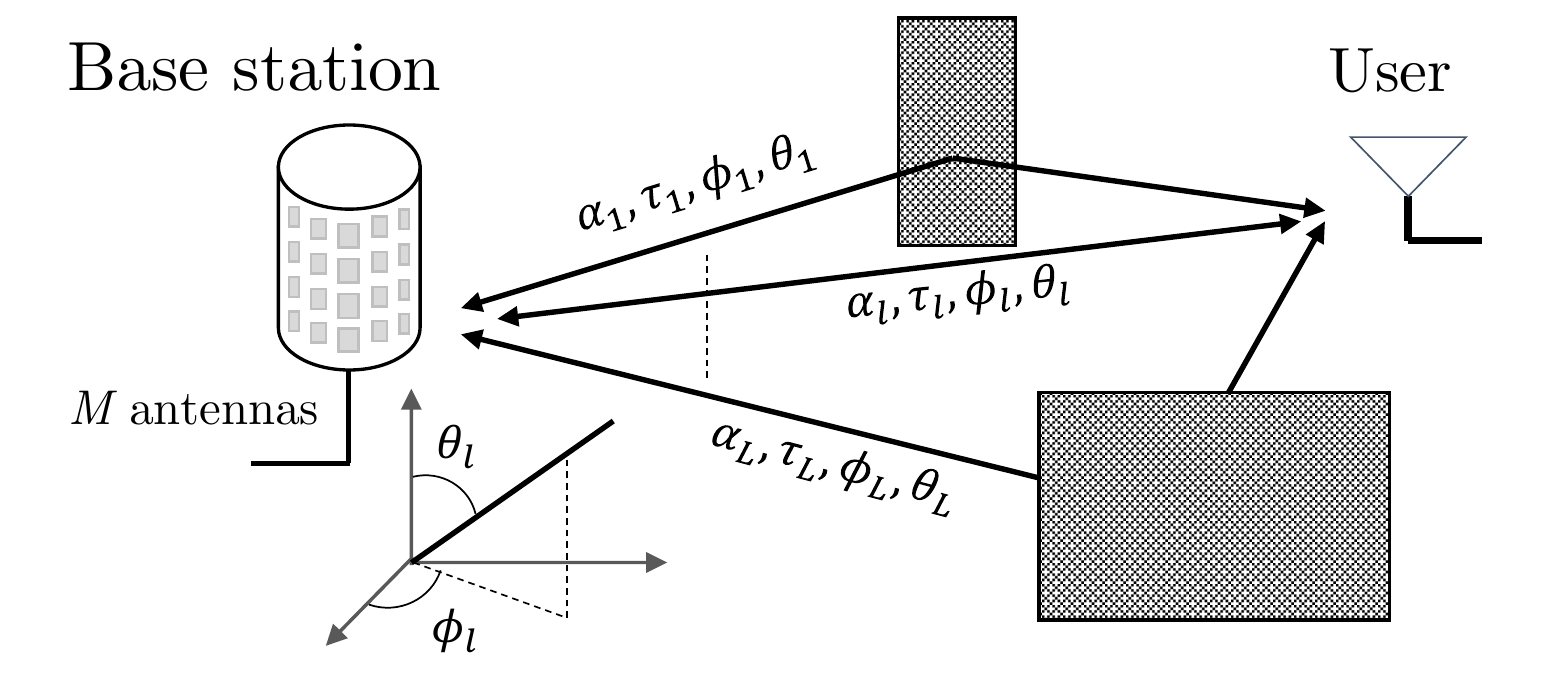}} 
	}
	\caption{Massive MIMO multipath propagation environment described by $L$ components. Each multipath component is characterized by its complex gain $\alpha_l$, its delay $\tau_l$, its azimuth angle $\phi_l$ and elevation angle $\theta_l$.}
	\label{fig:channel_model}
\end{figure}

Furthermore, we assume that the propagation channel is composed of $L$ specular paths, where each path is completely characterized by its deterministic parameters: complex gain $\alpha_l=\Re(\alpha_l)+\jmath \Im(\alpha_l)$, delay $\tau_l$, azimuth angle $\phi_l$ and elevation angle $\theta_l$, as depicted in Fig.~\ref{fig:channel_model}. All antennas are assumed to be vertically polarized. Under these assumptions, the channel frequency response $h_m(f)$ can be expressed as
\begin{align}
h_m(f)\triangleq\sum_{l=1}^{L} \alpha_l {a}_m(\phi_l,\theta_l,f) e^{-\jmath 2\pi f \tau_l} \label{eq:frequency_channel_response},
\end{align}
where ${a}_m(\phi,\theta,f)$ denotes the pattern of antenna $m$ evaluated in the direction $(\phi,\theta)$ and at frequency $f$. Note that the frequency dependence of the array pattern cannot generally be omitted, depending on the size of the band $B$ and the targeted extrapolation frequency range. More specifically, the frequency selectivity of ${a}_m(\phi,\theta,f)$ comes from two contributions: (i) the frequency dependence of each individual antenna pattern and (ii) the frequency dependent phase shift across the antenna array elements (beam squinting). This dependence is often neglected in the literature when the ratio of the dimension of the array to the speed of light is much smaller than the inverse of the bandwidth of the signal.

Furthermore, a number of straightforward generalizations of the model in (\ref{eq:frequency_channel_response}) can be made: (i) V and H polarizations can be taken into account by representing the path amplitudes as $2 \times 1$ vectors and the array patterns as $2 \times 2$ polarimetric matrices. (ii) multiple antenna elements at the UEs can also be taken into account by considering their array pattern and angles of departures; (iii) scatterers in the nearfield can be described by replacing the plane wave model of each path by a spherical wave model, where the wavefront curvature of each path is now an additional parameter of the model. However, for ease of exposition, we use the simplified model of (\ref{eq:received_signal_FD}) in the remainder of this paper.

We finally note that a further requirement for using uplink CSI for downlink beamforming is reciprocity calibration \cite{ReciprocityCalibration1,ReciprocityCalibration4,ReciprocityCalibration5}, since upconverters and downconverters might have different transfer functions that have to be compensated for by suitable calibration. Since reciprocity calibration affects FDD and TDD systems in the same manner, we disregard it in the following derivations and simulations,\textit{ i.e.}, we assume that it is perfect. 

 \section{Channel Estimation and Extrapolation}
\label{section:channel_extrapolation}
In this section, we first review conventional low-resolution channel estimators such as LS and LMMSE. We then explain the general concept of high-resolution channel estimation and we detail the principle of the SAGE algorithm. If the frequency of interest $f$ of the channel estimate $\hat{h}_m(f)$ is inside the training band $f\in[-B/2,B/2]$, we refer to the estimation process as interpolation. Otherwise, if $f\notin[-B/2,B/2]$, we refer to it as extrapolation and $f$ is also referred to as the extrapolation range. One can note that the interpolation performance corresponds to the channel estimation performance of a downlink TDD performance since uplink and downlink bands are shared in TDD mode.

\subsection{Conventional Low-Resolution Estimation}

As depicted in Fig.~\ref{fig:system_model}, LS estimators perform a simple per-antenna estimation at each pilot subcarrier as
\begin{align}
\hat{h}_{\mathrm{LS},m}(f_k)&=\frac{r_m(f_k)}{s(f_k)}=h_m(f_k)+\frac{w_m(f_k)}{s(f_k)}, \label{eq:LS_estimate}
\end{align}
for $k=0,...,K-1$. Based on the $K$ channel estimates obtained at each pilot subcarrier, a linear method is generally used to obtain the channel at non-pilot subcarriers. We here propose to use a LMMSE estimator \cite{Edfors1998}. Denoting the vector containing the LS estimates by  $\hat{\vect{h}}_{\mathrm{LS},m}\triangleq(\hat{h}_{\mathrm{LS},m}(f_0),\hdots, \hat{h}_{\mathrm{LS},m}(f_{K-1}))^T$, the LMMSE estimate at frequency $f$ is given by 
\begin{align}
	\hat{h}_{\mathrm{LMMSE},m}(f)= \vect{p}_m^H(f)\hat{\vect{h}}_{\mathrm{LS},m}, \label{eq:LMMSE_estimate}
\end{align}
where the vector of coefficient $\vect{p}_m^H(f)$ is obtained by minimizing the MSE of the estimate. Assuming that the complex channel $h_m(f_k)$ has a zero mean, this gives
\begin{align*}
	\vect{p}_m^H(f)&=\vect{c}^H_{\mathrm{LS},m}(f)\mat{C}_{\mathrm{LS},m}^{-1}\\ \mat{C}_{\mathrm{LS},m}&\triangleq\mathcal{E}\left[\hat{\vect{h}}_{\mathrm{LS},m}\hat{\vect{h}}_{\mathrm{LS},m}^H\right],\
	\vect{c}^H_{\mathrm{LS},m}(f)\triangleq \mathcal{E} \left[h_m(f) \hat{\vect{h}}^H_{\mathrm{LS},m}\right],
\end{align*}
where the expectation $\mathcal{E}[.]$ is taken over both the noise and channel statistics as opposed to $\mathbb{E}[.]$ which denotes the expectation with respect to the noise statistics only. We can further write
\begin{align*}
[\vect{c}^H_{\mathrm{LS},m}(f)]_k&=C_{h,m}(f,f_k)\\ [\mat{C}_{\mathrm{LS},m}]_{k,k'}&=C_{h,m}(f_k,f_{k'})+\frac{\sigma_w^2}{|s(f_k)|^2} \delta_{k-k'},
\end{align*}
where
\begin{align}
C_{h,m}(f,f')\triangleq \mathcal{E}\left(h_m(f)h_m(f')^*\right) \label{eq:def_frequ_autocorrelation_channel}
\end{align}
is the autocorrelation function of the channel frequency response. An implementation challenge of the LMMSE estimator is the computation of $C_{h,m}(f,f')$. This function depends on the joint distribution of the path parameters. In the following, we approximate the computation of $C_{h,m}(f,f')$ by assuming that, as in \cite{Edfors1998}, the paths gains and delays are i.i.d. We also assume a frequency independent pattern\footnote{This assumption is not very restrictive here given that, for typical values of $B$, the array can be considered frequency independent inside the training band $B$.} $a_m(\phi,\theta,f)=a_m(\phi,\theta)$ and isotropic array pattern $|a_m(\phi,\theta)|^2=1$. Furthermore, the delay of each path $\tau_l$ has a uniform distribution in $[0,\ \tau_{\mathrm{max}}]$ while the complex path gain $\alpha_l$ has a uniform power across delay. This gives
\begin{align*}
	C_{h,m}(f,f')
	&=\frac{L}{\tau_{\mathrm{max}}} \mathcal{E}\left[ |\alpha_l|^2  \right] \int_{0}^{\tau_{\mathrm{max}}} e^{-\jmath 2\pi  \Delta f\tau} d\tau\\
&= P_h e^{-\jmath \pi  \Delta f\tau_{\mathrm{max}}} \text{sinc} (\pi \Delta f \tau_{\mathrm{max}}),
\end{align*}
where $\Delta f=f-f'$ and $P_h=C_{h,m}(0)$ is the averaged channel power. The LMMSE estimator performs relatively well in-band as long as the pilot spacing is smaller than $1/\tau_{\mathrm{max}}$, in accordance with the Nyquist sampling theorem. However, its extrapolation performance degrades quickly out of the training band, as will be analytically studied in Section~\ref{section:Performance_limits}. An intuitive way to see this is to simply notice that the autocorrelation function $C_{h,m}(\Delta f)$ decays in $1/(\Delta f \tau_{\mathrm{max}})$. This implies that the extrapolation performance can only be satisfactory for extrapolation range $f$ spaced about ${1}/\tau_{\mathrm{max}}$ away from the training band, \textit{i.e.}, $f \in [-\frac{B}{2}-\frac{1}{\tau_{\mathrm{max}}},\frac{B}{2}+\frac{1}{\tau_{\mathrm{max}}}]$. For a typical delay spread of $\tau_{\mathrm{max}}=2.5\ \mu$s, we have $\frac{1}{\tau_{\mathrm{max}}}=400$ kHz, which is much too low to deploy a typical FDD massive MIMO system.

Note that, to compute the expectation $\mathcal{E}[.]$, we used the long term statistics of the path parameters. Following the same idea, the LMMSE estimator presented here did not combine the LS samples from different antennas and hence did not leverage spatial correlation to improve the performance. This is justified for the case that the angular spread is sufficiently large (for a given antenna spacing) such that the correlation between antennas is close to zero, \textit{i.e.}, $\mathcal{E}\left(h_m(f)h_{m'}(f)^*\right)\approx 0$ if $m\neq m'$; \textit{e.g.}, if the antenna spacing is half a wavelength and the angular distribution of the scatterers is uniform \cite{molisch2012wireless}. This implies that combining the LS estimates from different antennas would not provide any significant gain. 

On the other hand, an improved LMMSE estimator implementation would require to estimate the second-order statistics of the channel or ``instantaneous" distribution of the path parameters. Under typical propagation conditions, paths delays and angles are clustered and thus far from being uniformly distributed. This implies that the frequency-space correlation function can be first estimated and then leveraged to significantly improve the LMMSE performance \cite{Adhikary2013,Barzegar2019}. However, this gain is properly taking into account by the high-resolution estimator proposed in the following.

\subsection{High-Resolution Estimation}
\label{section:high_resolution_motivation}

\begin{figure*}[!t]  
	\centering
	
	\resizebox{0.7\textwidth}{!}{%
		{\includegraphics[clip, trim=0cm 6.5cm 14cm 0cm, scale=1]{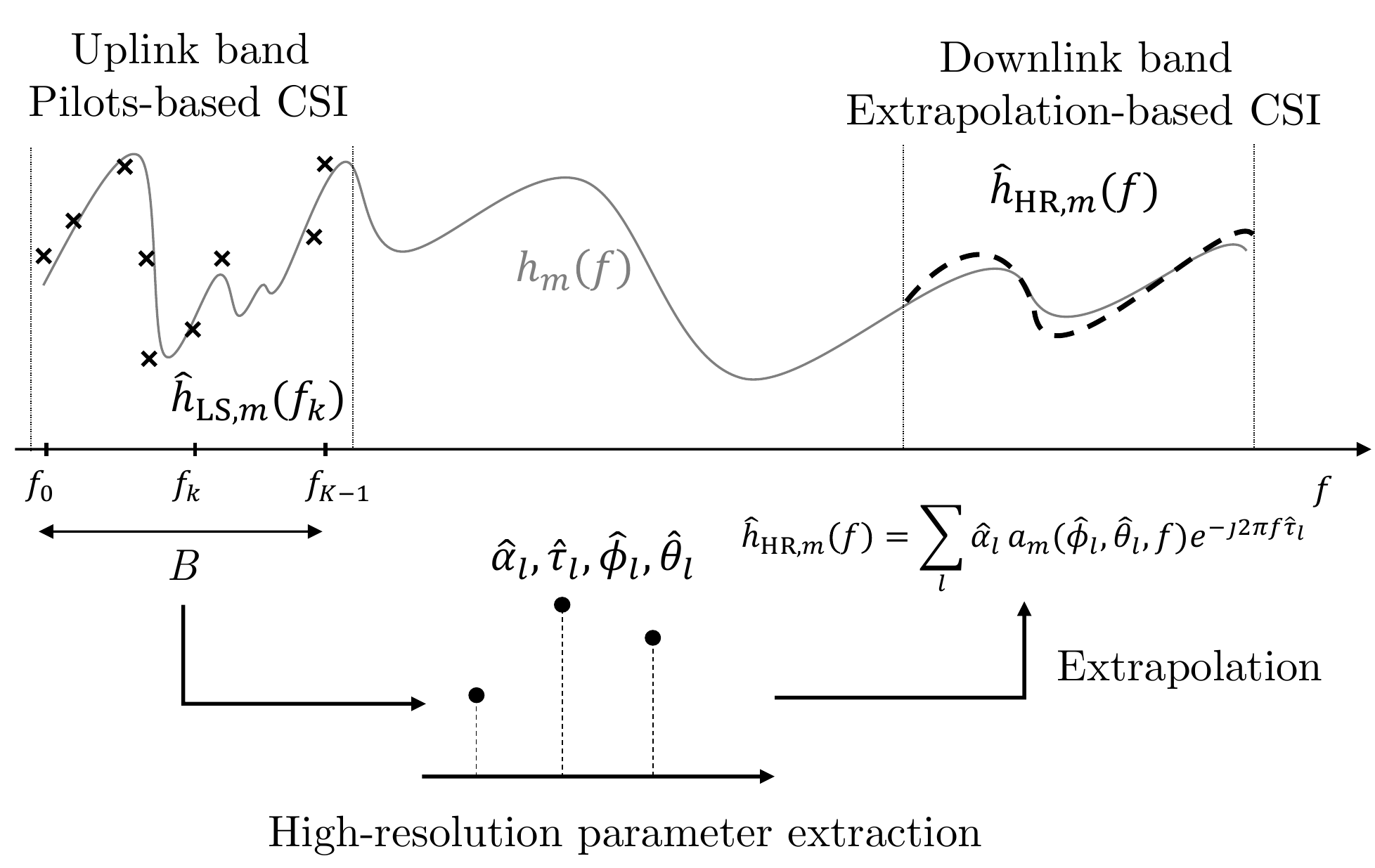}} 
	}
	\caption{Extrapolation-based FDD system: channel in the downlink band is extrapolated from pilots received in the uplink band, relying on high-resolution parameter extraction. CSI: channel state information.}
	\label{fig:HR_estimation}
\end{figure*}

The poor extrapolation performance of the LMMSE estimator can be intuitively explained by the fact that the extrapolated channel does not exhibit a linear dependence on the in-band channel frequency response. On the other hand, high-resolution channel estimation allows to alleviate these limitations. As depicted in Fig.~\ref{fig:HR_estimation}, instead of estimating the composite channel function $h_m(f)$, the HRPE approach directly estimates the parameters of each path. Taking advantage of the underlying (non-linear) physical dependence of the channel on its parameters can prove very useful to improve the extrapolation performance.

If we denote by $\hat{\tau}_l$, $\hat{\phi}_l$, $\hat{\theta}_l$ and $\hat{\alpha}_l$ the high-resolution estimates of ${\tau}_l$, ${\phi}_l$, ${\theta}_l$ and ${\alpha}_l$ respectively, the high-resolution (HR) estimate of the extrapolated channel reads as
\begin{align}
\hat{h}_{\mathrm{HR},m}(f)=\sum_{l=1}^{L} \hat{\alpha}_l {a}_m(\hat{\phi}_l,\hat{\theta}_l,f) e^{-\jmath 2\pi f \hat{\tau}_l} \label{eq:channel_frequency_extrapolated}.
\end{align}
{\color{black} Of course, intuitive reasoning tells us that the extrapolated channel will suffer from the estimation errors on the path parameters. Indeed, the finite bandwidth and aperture of the array directly induce a finite resolution in delay and angle, which leads to inaccuracy in the estimation of the delay and angle parameters. Moreover, the error on $\hat{h}_{\mathrm{HR},m}(f)$ becomes especially large as the extrapolation range becomes large. Indeed, one can note that the delay estimates $\hat{\tau}_l$ are multiplied by $f$ so that the error increases as $f$ increases. The impact of inaccurate parameter estimation will be carefully studied in Section~\ref{subsection:HRE}.} We also assume here that the parameters of the MPCs are independent of frequency. In other words, the channel is assumed to remain stationary on a frequency band including both uplink and downlink FDD disjoint bands. This is well fulfilled in most practical situations \cite{Molisch2009}, since these parameters remain constant over a bandwidth corresponding to about $10\%$ of the carrier frequency. For sub-6 GHz systems, this assumption is generally satisfied so that channel extrapolation could be used. Indeed, the LTE duplex spacing between uplink and downlink bands is: 10 to 70 MHz in the 700-900 MHz band and 45 to 190 MHz in the 1400-1700 MHz band. An exception are the Advanced Wireless Services (AWS) bands of LTE, which have up to 400 MHz spacing at 1700-2100 MHz carrier frequency.

To extract the path parameters $\hat{\tau}_l$, $\hat{\phi}_l$, $\hat{\theta}_l$ and $\hat{\alpha}_l$, we propose to use the SAGE algorithm originally described in \cite{Fleury1999} and widely adopted in the channel propagation community. We extended the algorithm to extract elevation angles and take into account the frequency dependence of the array. One should note that the main contribution of this paper is not to propose a novel efficient algorithm for high-resolution channel estimation but rather to show that the general performance bounds derived in the next section can be approached by conventional algorithms with reasonable complexity, such as the SAGE algorithm. Other HRPE algorithms could also be applied and may provide different results.

In brief, the algorithm aims at maximizing the likelihood of the received samples at pilot subcarriers as a function of the path parameters. Let us define $\vect{\psi}=(\vect{\psi}_1^T,\hdots,\vect{\psi}_L^T)^T \in \mathbb{R}^{5L\times 1}$ and $\vect{\psi}_l=({\tau}_l, {\phi}_{l}, {\theta}_{l}, \Re({\alpha}_l), \Im ({\alpha}_l))^T \in \mathbb{R}^{5\times 1}$ as the vectors containing all real path parameters and the real parameters of each path respectively. The vectors $\hat{\vect{\psi}}$ and $\hat{\vect{\psi}}_l$ respectively denote their estimate. The optimization problem of maximizing the likelihood can be reformulated as
\begin{align*}
 \min_{\hat{\vect{\psi}}} \sum_{m=1}^M\sum_{k=0}^{K-1} \left| r_m(f_k) - \sum_{l=1}^{L} \hat{\alpha}_l {a}_m(\hat{\phi}_l,\hat{\theta}_l,f) e^{-\jmath 2\pi f \hat{\tau}_l} s(f_k) \right|^2.
\end{align*}
The problem is not easy to solve due to its high-dimensionality ($5L$) and the highly non-linear dependence on the path parameters. The SAGE algorithm provides an efficient suboptimal solution to the problem relying on an iterative approach. At each iteration, only the parameters corresponding to one path, \textit{e.g.}, $\vect{\psi}_l$, are optimized while other path parameters keep their past value. This reduces the search dimensions from $5L$ dimensions to $5$ dimensions at each iteration. Furthermore, inside each iteration, the 5-dimensional search is simplified into five one-dimensional searches optimizing each parameter one at a time using a line search. The algorithm iterates until convergence or if a maximal number of iterations is achieved. The initial estimates of each path are obtained by successive ordered cancellation.

\section{Performance Analysis}
\label{section:Performance_limits}

In this section, we start by studying the performance of the previously detailed algorithms in terms of the MSE of the estimated channel frequency response and the related error correlation matrix. In the last part of this section, we study the relationship between the MSE and the downlink user performance. More specifically, we will derive the expression of the SNR at the user side, taking into account the beamforming power loss induced by incorrect channel estimates.

We define the MSE of an estimate $\hat{h}_m(f)$ of $h_m(f)$ as
\begin{align}
\mathrm{MSE}_m(f)&\triangleq \mathbb{E} \left[|\hat{h}_m(f)-{h}_m(f)|^2\right], \label{eq:def_MSE}
\end{align}
where the expectation is taken over the noise realizations for a fixed channel realization ${h}_m(f)$ and thus the underlying parameters $\vect{\psi}$. Similarly, the error correlation matrix of the channel vector at frequency $f$ is defined as
\begin{align*}
\mat{E}(f)&\triangleq \mathbb{E} \left[ \left(\vect{h}(f)-\hat{\vect{h}}(f)\right)\left(\vect{h}(f)-\hat{\vect{h}}(f)\right)^H  \right]	
\\
[\mat{E}(f)]_{m,m'}&=\mathbb{E}\left[\left(\hat{h}_{m}(f)-{h}_m(f)\right)\left(\hat{h}_{m'}(f)-{h}_{m'}(f)\right)\right],
\end{align*}
with $\mathrm{MSE}_m(f)=[\mat{E}(f)]_{m,m}$, $\vect{h}(f)=(h_1(f),...,h_M(f))^T$ and $\hat{\vect{h}}(f)=(\hat{h}_1(f),...,\hat{h}_M(f))^T$.

\subsection{Conventional Low-Resolution Estimation}

To simplify the following expressions of the performance of the LS and LMMSE estimators, we will assume equipowered pilot symbol, \textit{i.e.}, $|s(f_k)|^2=E_s$. This assumption is consistent with, \textit{e.g.}, the Zhadoff-Chu training sequences in LTE and NR. Extension to the general case is straightforward. The total training power is then $E_T=\sum_{k=0}^{K-1} |s(f_k)|^2=KE_s$. Based on the expression of the LS estimate in (\ref{eq:LS_estimate}), the MSE expression at pilot subcarrier $f_k$ is obtained as
\begin{align}
\mathrm{MSE}_{\mathrm{LS},m}(f_k)&= \mathbb{E}\left[|\hat{h}_{\mathrm{LS},m}(f_k)-{h}_m(f_k)|^2\right] 
=\frac{\sigma_w^2}{E_s}. \label{eq:MSE_LS}
\end{align}
Moreover, the channel estimation errors at different antennas are uncorrelated given that noise samples are uncorrelated implying that $\mat{E}_{\mathrm{LS}}(f_k)=\frac{\sigma_w^2}{E_s} \mat{I}_M$. 

Defining ${\vect{h}}_{m}\triangleq({h}_{m}(f_0),\hdots, {h}_{m}(f_{K-1}))^T$ and using the expression of the LMMSE estimate in (\ref{eq:LMMSE_estimate}), the LMMSE performance for any frequency $f$ in-band (interpolation) or out-of-band (extrapolation) is
\begin{align*}
	\mathrm{MSE}_{\mathrm{LMMSE},m}(f)
&= |h_m(f)|^2-2\Re\left( h_m^*(f) \vect{p}_m^H(f){\vect{h}}_{m} \right)\\
&+\vect{p}_m^H(f)    \left( {\vect{h}}_{m}{\vect{h}}_{m}^H +\frac{\sigma_w^2}{E_s}\mat{I}_K  \right)            \vect{p}_m(f)\\
[\mat{E}_{\mathrm{LMMSE}}(f)]_{m,m'}
&=h_m(f)h_{m'}^*(f)\\
&\hspace{-1.5em} +\vect{p}_m^H(f)    \left( {\vect{h}}_{m}{\vect{h}}_{m'}^H +\frac{\sigma_w^2}{E_s}\mat{I}_K \delta_{m-m'} \right)            \vect{p}_{m'}(f)\\
&- h_m(f) {\vect{h}}_{m'}^H  \vect{p}_{m'}(f) - \vect{p}_{m}^H(f) {\vect{h}}_{m}  h_{m'}^*(f).
\end{align*}
Note that we took the expectation only over the noise statistics and not the channel statistics ($\mathbb{E}(.)$ instead of $\mathcal{E}(.)$). Under these statistics, the LMMSE estimator is biased, \textit{i.e.},  $\mathbb{E}\left[\hat{h}_{\mathrm{LMMSE},m}(f)\right]= \vect{p}_m^H(f){\vect{h}}_{m}\neq h_m(f)$.

\subsection{High-Resolution Estimation}
\label{subsection:HRE}
{\color{black}The MSE performance of high-resolution estimation heavily depends on the choice of the algorithm and the result is typically not in closed-form. To circumvent this limitation and make our result more general and tractable, we will compute the CRLB of the channel estimate, which is by definition a theoretical bound and is independent of the choice of the algorithm. In other words, the goal of this paper is not to derive specific channel extrapolation algorithms and to study their computational complexity, but we propose general theoretical performance bounds. In the simulation section, we will show that SAGE performs close to the CRLB, implying that the bound can be approached by conventional algorithms and hence is useful. }

To derive this bound, we can first notice that $\vect{h}(f)$ is a non-linear function of path parameters $\vect{\psi}$, as explicitly detailed in (\ref{eq:frequency_channel_response}). Using this fact, we can apply the CRLB formula for non-linear transformation of parameters \cite{kay1993fundamentals}. The bound tells us that for any unbiased estimator $\hat{\vect{h}}(f)$ of $\vect{h}(f)$, we have
\begin{align}
	\mat{E}(f)\succcurlyeq \mat{C}(f)\triangleq \left(\mat{G}(f)\right)^H\mat{I}_{\psi}^{-1}\mat{G}(f), \label{eq:CRLB_covariance_matrix}
\end{align}
where matrices $\mat{G}(f)$ and $\mat{I}_{\psi}$ are the Jacobian and Fisher information matrices respectively, whose forms are given in following subsections. The relationship $\mat{E}(f)\succcurlyeq \mat{C}(f)$ implies that the matrix $\mat{E}(f)-\mat{C}(f)$ is positive semidefinite, which directly implies that the MSE at antenna $m$ and frequency $f$ can be bounded by the corresponding diagonal element
\begin{align}
	\mathrm{MSE}_{m}(f)\geq [\mat{C}(f)]_{m,m}=\mat{g}_{m,f}^H\mat{I}_{\psi}^{-1} \mat{g}_{m,f}, \label{eq:CRLB_MSE}
\end{align}
where $\mat{g}_{m,f}$ is the $m$-th column of $\mat{G}(f)$.

\subsubsection{Fisher information matrix}
Matrix $\mat{I}_{\psi}\in \mathbb{R}^{5L\times 5L}$ is the Fisher information matrix of the path parameters. Since the received samples ${r}_m(f_k)$ at each antenna and pilot subcarrier follow a circularly symmetric complex normal distribution with variance $\sigma_w^2$ and mean
\begin{align*}
\mu_{m,k}\triangleq\sum_{l=1}^{L}{{\alpha}_l}  {a}_m(\phi_l,\theta_l,f_k)   e^{-\jmath 2\pi f_k \tau_l} s(f_k),
\end{align*}
we can directly use the CRLB formula for the general Gaussian case \cite{kay1993fundamentals} to compute each element of the Fisher information matrix
\begin{align}
\left[\mat{I}_{\vect{\psi}}\right]_{u,v}&=\frac{2}{\sigma_w^2}\sum_{k=0}^{K-1}\sum_{m=1}^{M} \Re \left( \frac{\partial \mu_{m,k}^*}{\partial \psi_u} \frac{\partial \mu_{m,k}}{\partial \psi_v} \right). \label{eq:Fisher_info_matrix_element}
\end{align}

The full $5L\times 5L$ Fisher information matrix $\mat{I}_{\vect{\psi}}$ can be partitioned into $L^2$ submatrices $\mat{I}_{\vect{\psi}_l,\vect{\psi}_{l'}}\in \mathbb{R}^{5\times 5}$ as
\begin{align}
\mat{I}_{\vect{\psi}}&=\frac{2}{\sigma_w^2}\begin{pmatrix}
\mat{I}_{\vect{\psi}_1,\vect{\psi}_1}& \hdots & \mat{I}_{\vect{\psi}_1,\vect{\psi}_L}\\
\vdots& \ddots & \vdots\\
\mat{I}_{\vect{\psi}_L,\vect{\psi}_1}& \hdots & \mat{I}_{\vect{\psi}_L,\vect{\psi}_L}
\end{pmatrix}\label{eq:Fisher_information_matrix},\\
\mat{I}_{\vect{\psi}_l,\vect{\psi}_{l'}}&=\begin{pmatrix}
I_{\tau_l \tau_{l'}}       & I_{\tau_l \phi_{l'}}       & I_{\tau_l \theta_{l'}}        & I_{\tau_l \alpha^R_{l'}}       & I_{\tau_l \alpha^I_{l'}}\\
I_{\tau_l \phi_{l'}}     & I_{\phi_l \phi_{l'}}       & I_{\phi_l \theta_{l'}}        & I_{\phi_l \alpha^R_{l'}}       & I_{\phi_l \alpha^I_{l'}}\\
I_{\tau_l \theta_{l'}}   & I_{\phi_l \theta_{l'}}   & I_{\theta \theta_{l'}}      & I_{\theta \alpha^R_{l'}}     & I_{\theta \alpha^I_{l'}}\\
I_{\tau_l \alpha^R_{l'}} & I_{\phi_l \alpha^R_{l'}} & I_{\theta \alpha^R_{l'}}  & I_{\alpha^R_l \alpha^R_{l'}}   & I_{\alpha^R_l \alpha^I_{l'}}\\
I_{\tau_l \alpha^I_{l'}} & I_{\phi_l \alpha^I_{l'}} & I_{\theta \alpha^I_{l'}}  & I_{\alpha^R_l \alpha^I_{l'}} & I_{\alpha^I_l \alpha^I_{l'}}
\end{pmatrix}.\nonumber
\end{align}
Defining $\dot{a}_{m,\phi}(\phi,\theta,f)\triangleq\frac{d a_m(\phi,\theta,f)}{d \phi}$ and $\dot{a}_{m,\theta}(\phi,\theta,f)\triangleq\frac{d a_m(\phi,\theta,f)}{d \theta}$, we can write the partial derivatives appearing in (\ref{eq:Fisher_info_matrix_element}) as
\begin{align*}
\frac{d \mu_{m,k}}{d \tau_l}&= \alpha_l{a}_m(\phi_l,\theta_l,f_k)   (-\jmath 2\pi f_k)s(f_k)  e^{-\jmath 2\pi f_k \tau_l} \\
\frac{d \mu_{m,k}}{d \phi_l}&= \alpha_l \dot{a}_{m,\phi}(\phi,\theta,f_k)  s(f_k)  e^{-\jmath 2\pi f_k \tau_l} \\
\frac{d \mu_{m,k}}{d \theta_l}&= \alpha_l \dot{a}_{m,\theta}(\phi,\theta,f_k)  s(f_k)  e^{-\jmath 2\pi f_k \tau_l}\\
\frac{d \mu_{m,k}}{d \alpha^R_l}&= {a}_m(\phi_l,\theta_l,f_k)   s(f_k)  e^{-\jmath 2\pi f_k \tau_l}\\
\frac{d \mu_{m,k}}{d \alpha^I_l}&= \jmath{a}_m(\phi_l,\theta_l,f_k)   s(f_k)  e^{-\jmath 2\pi f_k \tau_l}.
\end{align*}
Inserting these partial derivatives in (\ref{eq:Fisher_info_matrix_element}) and for a specific array pattern $a_m(\phi,\theta,f)$, the Fisher information matrix in (\ref{eq:Fisher_information_matrix}) can be easily constructed. In the following, we will make the assumption.

$\mathbf{(As1)}$: the Fisher information matrix $\mat{I}_{\vect{\psi}}$ is nonsingular.

In practice, a rank deficiency of $\mat{I}_{\vect{\psi}}$ could arise if several paths become close in delay and angle, which would cause the determinant of $\mat{I}_{\vect{\psi}}$ to go to zero. To be more accurate, the definition of ``close distance in delay and angle" should always be measured relatively to the system Fourier resolution. For instance, if the system occupies a 10 MHz bandwidth ($B=10$ MHz), inducing a resolution in delay of $1/B=100$ ns, two rays are said to be close in delay if their spacing is much smaller than 100 ns. Depending on the underlying physical phenomenon inducing the presence of dense multipath components, different solutions may be possible to address a rank deficiency of $\mat{I}_{\vect{\psi}}$.

If two rays, or more, are closely spaced in angle and delay and their delay separation is not only smaller than $1/B$ but also much smaller than $1/f$ where $f$ is the targeted extrapolation range, then they can be replaced by a single ray whose complex gain is given by the sum of the complex amplitudes of the correlated rays. As an example, if the targeted extrapolation performance is $f=100$ MHz, two rays with a spacing smaller than $1/f=10$ ns can be combined without affecting the extrapolation performance significantly.

If the source of dense multipath components is related to wavefront curvature, a more advanced channel model can be taken into account to address them \cite{Yin2017,le2018massive}. Indeed, our channel model in (\ref{eq:frequency_channel_response}) relies on a plane wave assumption. The presence of spherical waves would result in a large number of plane waves in (\ref{eq:frequency_channel_response}), inducing a potential ill-conditioning of $\mat{I}_{\vect{\psi}}$.

If, finally, the dense multipath components are present due to a truly rich scattering environment, instead of trying to estimate a large number of possibly unreliable paths, a better solution may be to consider them as random components. Rather than estimating their instantaneous values, their statistics can be estimated and included in the likelihood formulation to improve the conditioning of $\mat{I}_{\vect{\psi}}$. This approach is used by, \textit{e.g.}, the RIMAX algorithm \cite{richter2005estimation}.

\subsubsection{Jacobian matrix}
Matrix $\mat{G}(f) \in \mathbb{C}^{5L \times M}$ is the Jacobian matrix of the transformation defined as
\begin{align*}
	\mat{G}(f)&\triangleq \frac{\partial \vect{h}^T(f)}{\partial \vect{\psi}},\ [\mat{G}(f)]_{v,m}=\frac{\partial {h}_m(f)}{\partial {\psi}_v}.
\end{align*}
It can be partitioned into columns corresponding to each antenna element as $\mat{G}(f)=(\vect{g}_{1,f},...,\vect{g}_{M,f})$. Furthermore, each vector $\vect{g}_{m,f}$ can be partitioned into different paths and path parameters as
\begin{align*}
\vect{g}_{m,f}&= 
(\vect{g}_{m,f,\vect{\psi}_1}^T,\hdots,\vect{g}_{m,f,\vect{\psi}_L}^T)^T\\
\vect{g}_{m,f,\vect{\psi}_l}&= 
({g}_{m,f,\tau_l},{g}_{m,f,\phi_l},{g}_{m,f,\theta_l},{g}_{m,f,\alpha^I_l},{g}_{m,f,\alpha^R_l})^T\\
{g}_{m,f,\tau_l}&=(-\jmath 2\pi f) \alpha_l {a}_m({\phi}_l,{\theta}_l,f)  e^{-\jmath 2\pi f \tau_{l}}\\
{g}_{m,f,\phi_l}&=\alpha_l \dot{a}_{m,\phi}({\phi}_l,{\theta}_l,f)  e^{-\jmath 2\pi f \tau_{l}}\\
{g}_{m,f,\theta_l}&=\alpha_l \dot{a}_{m,\theta}({\phi}_l,{\theta}_l,f)  e^{-\jmath 2\pi f \tau_{l}}\\
{g}_{m,f,\alpha^R_l}&={a}_m({\phi}_l,{\theta}_l,f)  e^{-\jmath 2\pi f \tau_{l}}\\
{g}_{m,f,\alpha^I_l}&=\jmath {a}_m({\phi}_l,{\theta}_l,f)  e^{-\jmath 2\pi f \tau_{l}}.
\end{align*}

\subsubsection{Separated Rays}

The CRLB of (\ref{eq:CRLB_MSE}) is in closed-form, which allows to easily evaluate it numerically. However, it requires the inversion of the Fisher information matrix and does not provide much intuition on the extrapolation range that can be expected. To further characterize and try to gain more insight, let us introduce the set of assumptions $\mathbf{(As2)-(As4)}$.

$\mathbf{(As2)}$: the array pattern is non frequency selective, \textit{i.e.}, $a_m(\phi,\theta,f)=a_m(\phi,\theta)$.

This assumption does not generally depend on the channel but rather on the system parameters such as type of BS antennas, extrapolation range and carrier frequency. The assumption particularly makes sense if the antenna patterns are flat in the considered band and if the ratio of the dimension of the array to the speed of light is much smaller than that the inverse of the extrapolation range.

In the remaining part of this section, we assume that $\mathbf{(As2)}$ holds, and we drop the frequency dependence of the array. We define the following vectors in order to introduce assumptions $\mathbf{(As3)-(As4)}$
\begin{align*}
&\vect{s}_{l}\triangleq\begin{pmatrix}
s(f_0)e^{-\jmath 2\pi f_0 \tau_l}& \hdots & s(f_{K-1})e^{-\jmath 2\pi f_{K-1} \tau_l}
\end{pmatrix}^T\\
&\dot{\vect{s}}_{l}\triangleq -\jmath 2\pi (f_0 s(f_0)e^{-\jmath 2\pi f_0 \tau_l},...,f_{K-1}  s(f_{K-1})e^{-\jmath 2\pi f_{K-1} \tau_l})^T\\
&\vect{a}_l\triangleq\begin{pmatrix}
{a}_1(\phi_{l},\theta_{l})& \hdots &{a}_M(\phi_{l},\theta_{l})
\end{pmatrix}^T\in \mathbb{C}^{M\times 1}\\
&\dot{\vect{a}}_{l,\phi}\triangleq\begin{pmatrix}
\dot{a}_{1,\phi}(\phi_l,\theta_l)& \hdots &\dot{a}_{M,\phi}(\phi_l,\theta_l)
\end{pmatrix}^T\in \mathbb{C}^{M\times 1}\\
&\dot{\vect{a}}_{l,\theta}\triangleq\begin{pmatrix}
\dot{a}_{1,\theta}(\phi_l,\theta_l) \hdots &\dot{a}_{M,\theta}(\phi_l,\theta_l)
\end{pmatrix}^T\in \mathbb{C}^{M\times 1}.
\end{align*}

$\mathbf{(As3)}$: separation of the $L$ specular rays in delay, azimuth angle and/or elevation angle. We assume that, for each pair of rays $l,l'$ ($l\neq l'$), at least one of the following two relationships is verified:

(1) Separation in delay:
\begin{align}
\vect{s}_{l}^H\vect{s}_{l'}=\dot{\vect{s}}_{l}^H\dot{\vect{s}}_{l'}=\dot{\vect{s}}_{l}^H{\vect{s}}_{l'}=0. \label{eq:separation_in_delay}
\end{align}

(2) Separation in azimuth and/or elevation angle:
\begin{align*}
\vect{a}_l^H\vect{a}_{l'}&=\dot{\vect{a}}_{l,\theta}^H\dot{\vect{a}}_{l',\theta}=\dot{\vect{a}}_{l,\phi}^H\dot{\vect{a}}_{l',\phi}=0\\
\dot{\vect{a}}_{l,\theta}^H\vect{a}_{l'}&=\dot{\vect{a}}_{l,\phi}^H\vect{a}_{l'}=\dot{\vect{a}}_{l,\phi}^H \dot{\vect{a}}_{l',\theta} =0.
\end{align*}

The assumption $\mathbf{(As3)}$ is a strong assumption, whose accuracy will typically depend on different parameters. The specular paths will generally become relatively more separated in delay as the bandwidth of $s(f)$ increases, inducing higher resolution in delay. Similarly, the resolution and hence the separation in azimuth and elevation will be improved as the number of antenna elements $M$ is increased. More generally, for a given channel, the validity of $\mathbf{(As3)}$ will depend on the training signal $s(f_k)$, on the array pattern $a_m(\phi,\theta)$ and on the extrapolation range. Moreover, the validity of $\mathbf{(As3)}$ will be assessed in Section~\ref{section:simulation_results} using practical channel models.

$\mathbf{(As4)}$: the transmitted pilots $s(f_k)$ have a symmetric energy distribution implying that $|s(f)|^2=|s(-f)|^2$ and
\begin{align}
\dot{\vect{s}}_{l}^H{\vect{s}}_{l}=(\jmath 2\pi) \sum_{k=0}^{K-1} f_k |s(f_k)|^2 =0,\quad l=1,\hdots,L. \label{eq:frequency_symmetry}
\end{align}
Furthermore, the array pattern $a_m(\phi,\theta)$ satisfies the following symmetry condition
\begin{align*}
\dot{\vect{a}}_{l,\phi}^H{\vect{a}}_{l}=\dot{\vect{a}}_{l,\theta}^H{\vect{a}}_{l}=0,\quad l=1,\hdots,L.
\end{align*}
The symmetric condition on the pilot energy is satisfied in conventional systems such as LTE or NR since pilots have uniform energy while the condition on the array pattern is generally satisfied for symmetric arrays. For instance, it is easy to check that the condition is fulfilled for a rectangular array if each antenna element has an isotropic pattern according to (\ref{eq:isotropic_pattern}) later studied in Section~\ref{section:simulation_results}. The following bound gives a particularization of the CRLB of (\ref{eq:CRLB_MSE}) under additional assumptions $\mathbf{(As2)}-\mathbf{(As4)}$ and for the MSE averaged over the receive antennas, \textit{i.e.},
\begin{align*}
\mathrm{MSE}(f)\triangleq \frac{1}{M}\tr\left[\mat{E}(f)\right]= \frac{1}{M}\sum_{m=1}^M\mathrm{MSE}_{m}(f).
\end{align*}

\begin{proposition} \label{proposition:well_separated_rays}
	Under $\mathbf{(As2)}-\mathbf{(As4)}$, the expression of the CRLB of (\ref{eq:CRLB_MSE}) averaged over the receive antennas simplifies to
	\begin{align}
	\mathrm{MSE}(f)&\geq \frac{\sigma_w^2}{E_T}\underbrace{\frac{L}{M}}_{\mathrm{SNR\ gain}}\left(\underbrace{2}_{\mathrm{Loss\ factor}}+\underbrace{\frac{1}{2}\left(\frac{f}{\sigma_F}\right)^2}_{\mathrm{Extrapolation\ penalty}} \right), \label{eq:simplified_LB}
	\end{align}
	where  $E_T$ is the total training power and $\sigma_F^2$ is the mean squared bandwidth of the transmit signal
	\begin{align*}
	E_T &= \|\vect{s}_l\|^2=\sum_{k=0}^{K-1} |s(f_k)|^2\\
	\sigma_F^2&\triangleq\frac{\|\dot{\vect{s}}_l\|^2}{(2\pi)^2\|\vect{s}_l\|^2}=\frac{\sum_{k=0}^{K-1} f_k^2 |s(f_k)|^2}{\sum_{k=0}^{K-1}  |s(f_k)|^2}.
	\end{align*}
	\begin{proof}
		The proof is given in the Appendix.
	\end{proof}
\end{proposition}

By adding some assumptions, the CRLB can be greatly simplified and provides much insight into the physical meaning of the different terms of the bound. We can clearly identify the two main advantages of high-resolution channel estimation. As compared to the LS estimation performance that we derived in (\ref{eq:MSE_LS}) where the total pilot power is $E_T=KE_s$, a gain of a factor $\frac{MK}{L}$ can be observed. This gain comes from two contributions: the array gain $M$ and the estimation of only $L$ channel coefficients instead of $K$ as in the LS case. However, a loss factor of 2 appears, coming from the penalty of estimating the azimuth and elevations angles of each path. {\color{black}Moreover, the channel can be extrapolated in frequency at the cost of a MSE penalty that quadratically scales with the ratio ${f}/{\sigma_F}$, which physically makes sense. Indeed, as the extrapolation range $f$ increases, the estimate quality worsens. On the other hand, as the uplink training bandwidth increases, the delays of each path are better estimated, which leads to an improved extrapolation performance. Note that the denominator $\sigma_F$ indicates that the extrapolation range can be quantified in multiples of the uplink training band $B$.}

It is interesting to see that the simplified CRLB does not depend on the path parameters $\vect{\psi}$ for well separated paths. This is in part explained by the fact that each path is well separated, which cancels the interdependence between different paths. Additionally, the channel frequency response is evaluated in the direction of the incoming specular waves, canceling the dependence on the parameters of each path as well as the dependence on the array pattern.

Based on the simplified CRLB, we can find a closed-form expression of the extrapolation range: we define the $\gamma$ extrapolation range, denoted by $f_{\mathrm{Extrapol-\gamma}}$, as the frequency $f$ beyond which the extrapolation performance falls $\gamma$ times below that of the conventional LS estimator given in (\ref{eq:MSE_LS}). Using the expressions of (\ref{eq:MSE_LS}) and (\ref{eq:simplified_LB}), we easily find
\begin{align}
f_{\mathrm{Extrapol-\gamma}}=2\sigma_F \sqrt{\frac{MK\gamma}{2L}-1}. \label{eq:extrapolation_range}
\end{align}
Note that this definition is independent of the ratio $E_s/\sigma_w^2$.

\subsection{Relationship between channel MSE and user SNR}
\label{subsection:Relation_MSE_SNR}
The goal of this section is to find how the user performance is affected by imperfect CSI at the BS. Let us assume that the BS communicates in the downlink by beamforming in the direction of the user using a beamforming vector $\vect{g}(f) \in \mathbb{C}^{M\times 1}$. The frequency $f$ denotes the pilot subcarrier frequency at which the symbol $d(f)$ is transmitted. The BS uses maximum ratio combining and normalizes the beamforming vector to have unit power so that 
\begin{align*}
	\vect{g}(f)=\frac{\left(\hat{\vect{h}}(f)\right)^*}{\|\hat{\vect{h}}(f)\|}.
\end{align*}
{\color{black} The demodulated symbol at the user side is
	\begin{align*}
	r(f)=(\vect{h}(f))^T \vect{g}(f) d(f) +w(f),
	\end{align*}
	and the related SNR, averaged over the statistics of the transmit symbols, the noise on uplink pilots and the noise on downlink symbols,
	\begin{align}
	\mathrm{SNR}_{\mathrm{DL}}(f)&=\frac{E_{d,f}}{\sigma_w^2} \mathbb{E}\left[|(\vect{h}(f))^T \vect{g}(f)|^2\right] \label{eq:SNR_DL} \\
	&= \frac{E_{d,f} \|\vect{h}(f)\|^2}{\sigma_w^2} \underbrace{\mathbb{E}\left[\frac{|(\hat{\vect{h}}(f))^H{\vect{h}}(f)|^2}{\|\hat{\vect{h}}(f)\|^2\|{\vect{h}}(f)\|^2}\right]}_{\eta(f)}, \nonumber
	\end{align}
	where the channel $\vect{h}(f)$ is considered as deterministic (not random) and $E_{d,f}$ is the energy of downlink symbol $d(f)$.} The term $\eta(f)$ is the so-called beamforming efficiency bounded as $0\leq \eta(f) \leq 1$. It represents the beamforming power loss due to imperfect CSI. If it close to 1, almost no loss is induced. On the other hand, if it is close to 0, the efficiency is strongly affected. {\color{black}We assume that coherent demodulation can be practically achieved at the user side by including pilots in the downlink transmission frame. This allows the users to estimate and compensate their equivalent channel after beamforming by the BS. Note that these pilots are beamformed to each user. They are thus orthogonal and can be transmitted at the same time-frequency resource to different users, implying that their overhead is negligible. Moreover, they do not need to be fed back to the BS. Note that the use of such user-specific reference signals for channel estimation is a common feature foreseen for 5G NR deployment \cite{Parkvall2017,dahlman20185g}. In the following, we neglect their overhead in the spectral efficiency computation. This is motivated by the fact that they are also present in TDD systems, implying that the comparison remains fair.}

{\color{black}The analytical expression of $\eta(f)$ is complicated to compute as it involves the expectation of a ratio of random variables. Therefore, we propose to approximate the expectation of the ratio by the ratio of expectations, which corresponds to the first-order Taylor expansion of the ratio around the mean of its numerator and denominator
\begin{align}
	\eta(f) \approx \hat{\eta}(f)&= \frac{\mathbb{E}\left[|(\hat{\vect{h}}(f))^H{\vect{h}}(f)|^2\right]}{\mathbb{E}\left[\|\hat{\vect{h}}(f)\|^2\|{\vect{h}}(f)\|^2\right]}\label{eq:approx_efficiency} \\
	&= \frac{\mathbb{E}\left[ |(\vect{h}(f)+\vect{e}(f))^H\vect{h}(f)|^2  \right] } {\mathbb{E}\left[\|\vect{h}(f)+\vect{e}(f)\|^2\right] \|{\vect{h}}(f)\|^2}, \nonumber
\end{align}
where we defined the estimation error $\vect{e}(f)\triangleq \hat{\vect{h}}(f)-\vect{h}(f)$. We expect the approximation $\hat{\eta}(f)$ to asymptotically converge to $\eta(f)$ as the number of antennas grows large. Indeed, one can note that both the numerator and the denominator of $\eta(f)$ involve a sum of $M$ elements. Thus, as $M$ grows large, the estimation error gets better averaged and we expect the numerator and denominator to converge to their expected value.} Based on the statistics of the estimation error $\vect{e}(f)$ derived in previous sections, $\hat{\eta}(f)$ can be easily computed. Furthermore, for an unbiased estimator, we have $\mathbb{E}(\vect{e}(f))=\vect{0}$ and $\hat{\eta}(f)$ simplifies to
\begin{align}
	\hat{\eta}(f)=\frac{\|\vect{h}(f)\|^2+\frac{(\vect{h}(f))^H \mat{E}(f)\vect{h}(f)}{\|\vect{h}(f)\|^2}}{\|\vect{h}(f)\|^2+\tr\left[\mat{E}(f)\right]}. \label{eq:efficiency_approx_simplified}
\end{align}
The second term of the numerator has the form of a Rayleigh quotient, which is upper and lower bounded by the maximal and minimal eigenvalues of matrix $\mat{E}(f)$ respectively. This implies that the numerator is always larger than the denominator, ensuring that $\hat{\eta}(f)\leq 1$ as expected.

{\color{black}In the end, we have found an analytical expression that relates the channel estimation error correlation matrix to the loss in beamforming power at the user side. The corresponding loss in terms of capacity, spectral efficiency and bit error rate can be directly inferred from the beamforming efficiency $\hat{\eta}(f)$. Indeed, the spectral efficiency at extrapolation range $f$ can be inferred as
\begin{align}
C(f)&=\log_2\left(1+\mathrm{SNR}_{\mathrm{DL}}(f) \right) \label{eq:spectral_efficiency}\\
&\approx \log_2\left(1+\frac{E_{d,f} \|\vect{h}(f)\|^2}{\sigma_w^2} \hat{\eta}(f) \right)\quad \text{[bits/symbols]} \nonumber.
\end{align}
Similarly, the uncoded symbol error rate for M-QAM symbols can be inferred as \cite{molisch2012wireless}
\begin{align}
	\text{SER}(f)&=2\frac{\sqrt{M}-1}{\sqrt{M}}\text{erfc}\left(\sqrt{\frac{3\mathrm{SNR}_{\mathrm{DL}}(f)}{2(M-1)}}\right)\label{eq:SER}\\
	&\approx 2\frac{\sqrt{M}-1}{\sqrt{M}}\text{erfc}\left(\sqrt{\frac{3E_{d,f} \|\vect{h}(f)\|^2 \hat{\eta}(f)}{2(M-1)\sigma_w^2} }\right), 	 \nonumber
\end{align}
where $\text{erfc}(.)$ is the complementary error function.}

Note that we only considered the performance of a single-user. The same methodology could be extended to study the performance of multiple users communicating at the same time and frequency. This type of study can be conducted relying on the formulas derived in this work for the statistics of the channel estimation errors and using mathematical tools from the random matrix theory literature \cite{Sanguinetti2019,Ozdogan2019}. Note that multi-user beamforming is more sensitive to channel estimation error as it can lead to inter-user interference.

\section{Simulation Results}
\label{section:simulation_results}

This section evaluates the performance of channel extrapolation for FDD massive MIMO system. The performance of the conventional LMMSE estimator will be compared to the high-resolution estimation based on the SAGE algorithm. The theoretical CRLB of the channel MSE derived in Section~\ref{subsection:HRE} is also included as a benchmark. Furthermore, the beamforming efficiency studied in Section~\ref{subsection:Relation_MSE_SNR} will be used to relate the MSE of the channel estimates to the user link performance in terms of SNR and spectral efficiency. Finally, graphs will include the performance of a corresponding TDD system. We emphasize that the loss of performance of the FDD versus TDD comes from the less accurate channel state information due to extrapolation. The TDD performance can be simply inferred from the previous analytical results as the performance related to the in-band channel estimation, which amounts to channel interpolation based on pilot estimates rather than extrapolation in the FDD mode.

\begin{figure*}[t]
	\centering 
	\subfloat[Channel estimate $\mathrm{MSE}(f)$.]{
		\resizebox{0.49\textwidth}{!}{
		\includegraphics{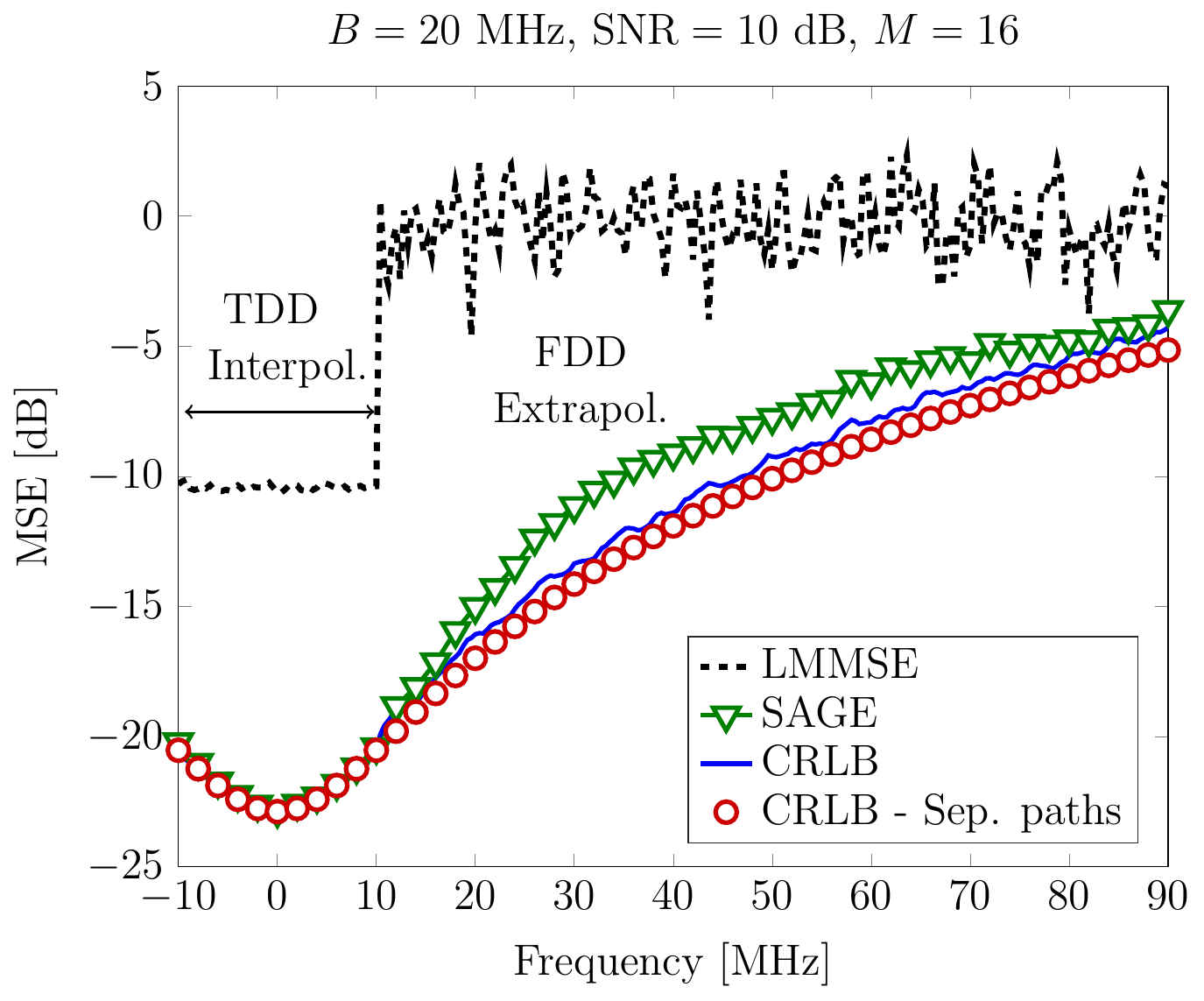}}
	}
	\subfloat[Beamforming efficiency $\eta(f)$.]{
		\resizebox{0.49\textwidth}{!}{
		\includegraphics{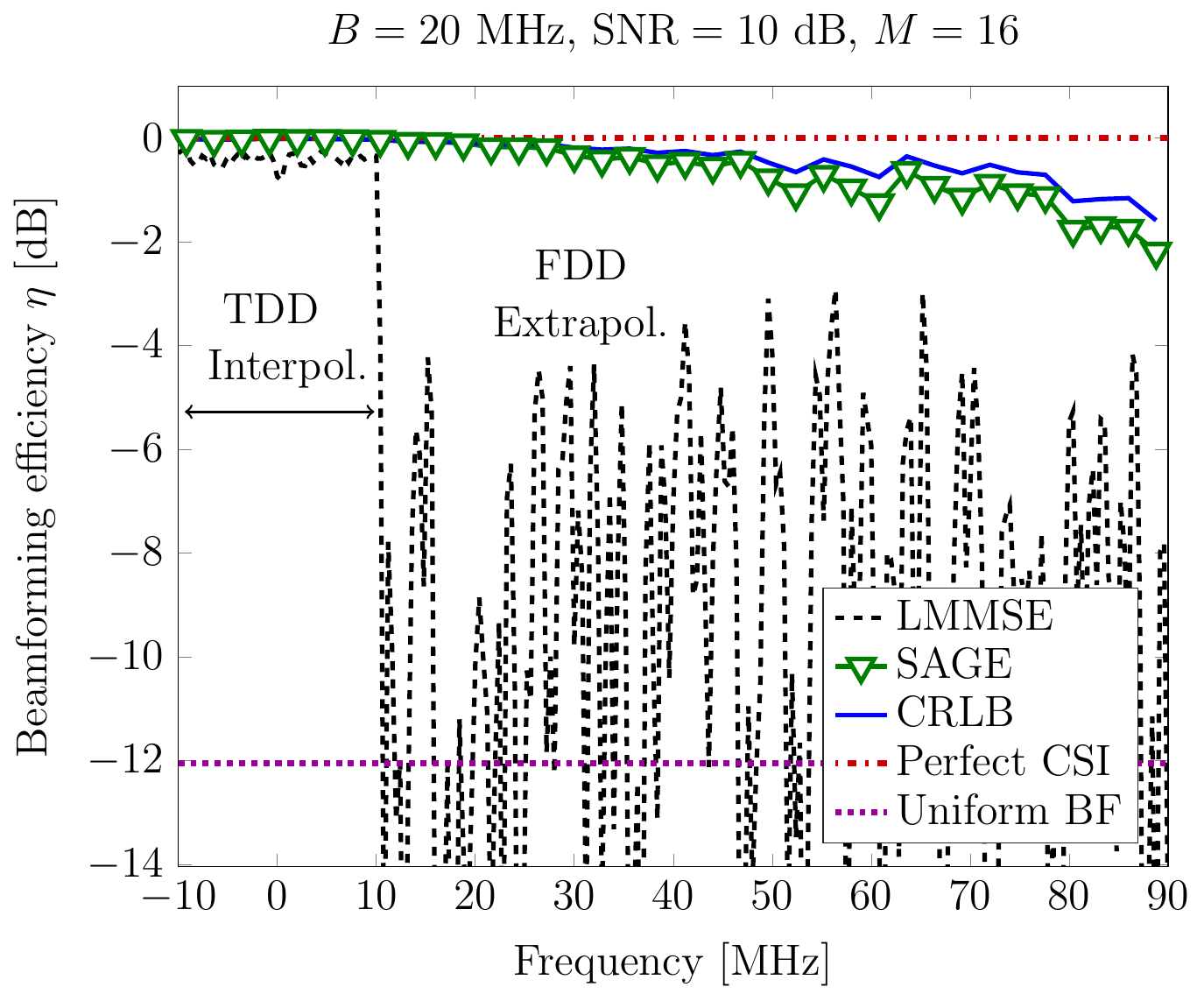}}
	}
	\caption{The LMMSE estimator performs poorly in terms of extrapolation while the high resolution SAGE estimator performs much better and approaches the CRLB. The simplified CRLB, assuming well separated paths, performs close to the CRLB.}
	\label{fig:SAGE_LMMSE} 
\end{figure*}

In the simulations, we assumed that the pilots $s(f_k)$ have uniform energy distribution over the $K$ frequency points $f_k$, \textit{i.e.}, $|s(f_k)|^2=E_s$ for $k=0,\hdots,K-1$. The pilots are uniformly spaced across the uplink bandwidth $B$ with spacing $1/\tau_{\mathrm{max}}$, \textit{i.e.}, $f_k=\left(k-\frac{K-1}{2}\right)/{\tau_{\mathrm{max}}}$ for $k=0,\hdots,K-1$ and $K=B\tau_{\mathrm{max}}+1$. This assumption is consistent with, \textit{ e.g.}, the Zhadoff-Chu training sequences in LTE and NR. We set $\tau_{\mathrm{max}}=2.5\mu s$. The center frequency of the uplink training band is set to $f_c=3.5$ GHz. Note that our system model is in baseband frequency. Hence, the carrier frequency $f_c$ corresponds to a zero baseband frequency ($f=0$). We consider a synthetic rectangular planar array at the BS with an inter-antenna element spacing of $\lambda_c/2$ where $\lambda_c$ is the center wavelength. The antenna elements have an isotropic pattern so that the pattern of each element becomes only a phase shift
\begin{align}
a_m(\phi,\theta,f)=e^{-\jmath 2\pi \frac{f_c + f}{c} \vect{r}_m\cdot \hat{\vect{e}}(\phi,\theta)}, \label{eq:isotropic_pattern}
\end{align}
where $\hat{\vect{e}}(\phi,\theta)$ is a unit vector in $\mathbb{R}^{3}$ pointing in the direction of the incoming ray $l$ and the position of the $m$-th receive array element is denoted by $\vect{r}_m \in \mathbb{R}^{3}$ with respect to a reference point. The reference point is chosen to ensure that $\sum_m \vect{r}_m =\vect{0}$. Note that (\ref{eq:isotropic_pattern}) is frequency dependent because of the beam squint effect. Three planar array geometries are considered: $M=4$ ($2\text{ Horiz.}\times 2\text{ Vert.}$), $M=16$ ($4\text{ Horiz.}\times 4\text{ Vert.}$) and $M=64$ ($8\text{ Horiz.}\times 8\text{ Vert.}$). 

The channel frequency response and received samples are generated according to (\ref{eq:frequency_channel_response}) and (\ref{eq:received_signal_FD}) respectively. The path parameters $\vect{\psi}$ are generated by the QuaDRiGa toolbox \cite{Jaeckel2014} according to the 3D-UMa NLOS model defined by 3GPP TR 36.873 v12.5.0 specifications \cite{3GPP_TR_36_873v12_5_0}. We took on purpose a non line-of-sight scenario to consider a more challenging case as all paths need to be resolved to properly model the channel instead of only a few in a line-of-sight case. The average channel power is normalized to one and the per-pilot SNR, defined as $\mathrm{SNR}\triangleq E_s/\sigma_w^2$, is set to 10 dB.


\subsection{High-Resolution versus LMMSE Estimation}


We start by analyzing the performance for a single realization of channel parameters. 
Fig.~\ref{fig:SAGE_LMMSE} compares the performance of the LMMSE and the high-resolution SAGE estimators and the system parameters $M=16$, $\mathrm{SNR}=10$ dB and $B=20$ MHz. This implies that the uplink pilots belong to the support $[-10,\ 10]$ MHz, which corresponds to the operating band of a corresponding TDD system since uplink and downlink bands are shared. The performance of the algorithms was averaged over 1000 noise realizations. A delay step size of $\frac{1}{50B}=1$ ns and an angular grid size of 1 degree are used as parameters of the SAGE grid search.

Fig.~\ref{fig:SAGE_LMMSE}. (a) depicts the performance in terms of the MSE of the channel estimates. The SAGE-based channel extrapolation approaches the MSE performance of the theoretical CRLB. This implies that the CRLB gives a good indication of the achievable MSE. Furthermore, we can see in the figure that the CRLB performs close to the simplified one of Proposition~\ref{proposition:well_separated_rays}, obtained under the assumption of well separated paths. {\color{black}On the other hand, the LMMSE estimator performs worse than high-resolution, especially out of the training band where the error appears to abruptly jump to higher values. Indeed, its transition zone is of the order of a few hundreds of kHz, which is negligible compared to the considered extrapolation range of about 100 MHz.} In-band, it also performs worse than SAGE as it does not exploit the joint spatial-frequency structure of the channel but performs independent per-antenna estimation.

Fig.~\ref{fig:SAGE_LMMSE}. (b) shows the beamforming efficiency $\eta(f)$ related to the different estimators. As a reminder, the beamforming efficiency, studied in Section~\ref{subsection:Relation_MSE_SNR}, corresponds to the loss of received signal power due to the impact of incorrect channel estimates on the beamformer. In the case of perfect CSI, $\eta(f)=1=0$ dB and the efficiency is maximized. Note that both estimators perform similarly in-band or in TDD mode. The reduction of $\eta(f)$ as the extrapolation frequency $f$ increases can be seen as the loss induced in the FDD system as compared to a TDD system. Here again, the LMMSE performance degrades very quickly out of the band. Note that its performance gets even worse at some points than a simple uniform beamforming strategy in all directions, \textit{i.e.}, $\vect{g}(f)=\vect{1}/\sqrt{M}$ implying a beamforming efficiency $\eta(f)=1/M\approx -12$ dB. On the other hand, the SAGE extrapolation approaches the efficiency related to the CRLB and only suffers from a beamforming power loss of about 2 dB at an extrapolation frequency of 90 MHz.

%

\subsection{Impact of Number of Antennas}

\begin{figure*}[t]
	\centering 
	\subfloat[ ]{
		\resizebox{0.49\textwidth}{!}{
		\includegraphics{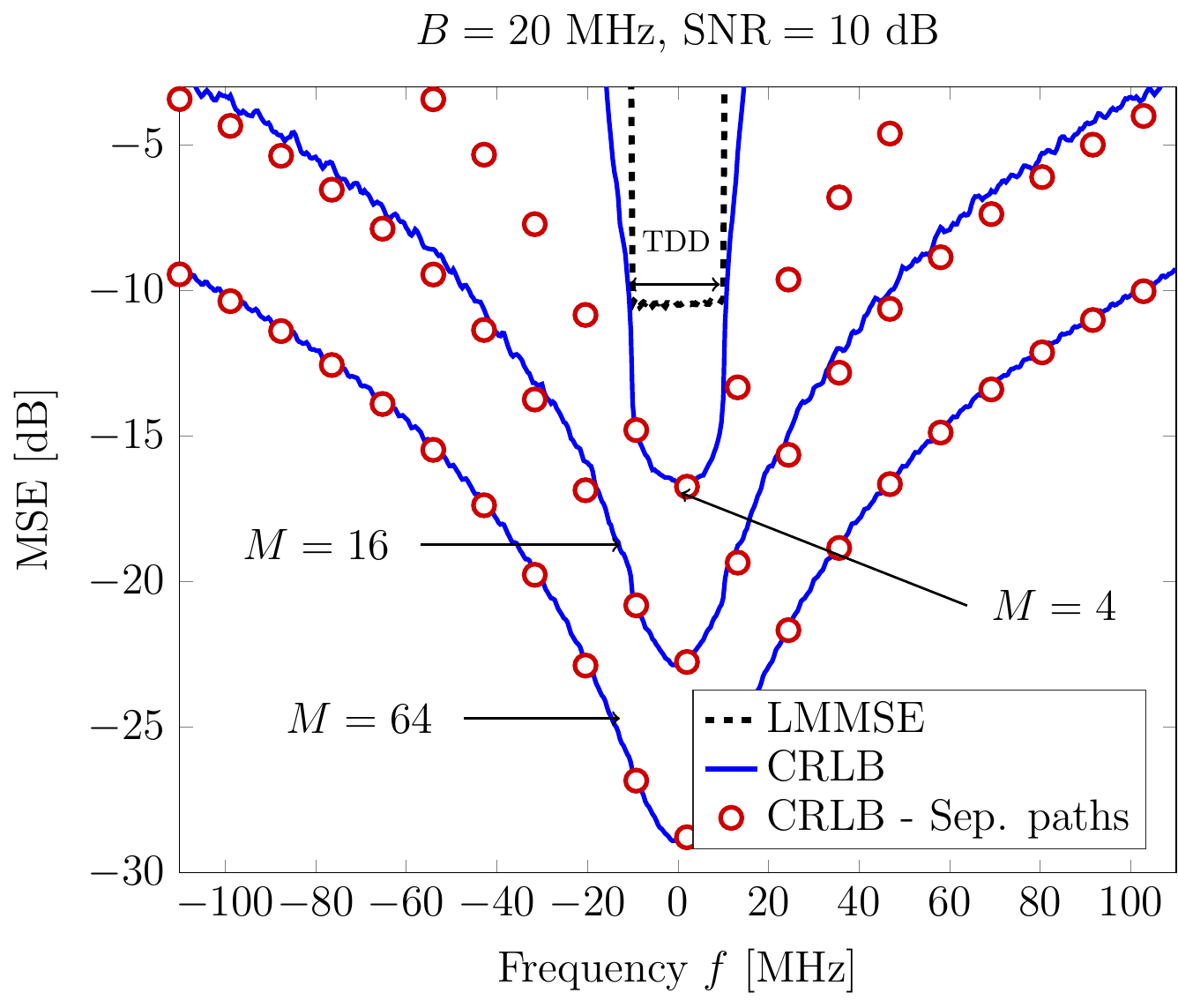}}
	}
	\subfloat[ ]{
		\resizebox{0.49\textwidth}{!}{
		\includegraphics{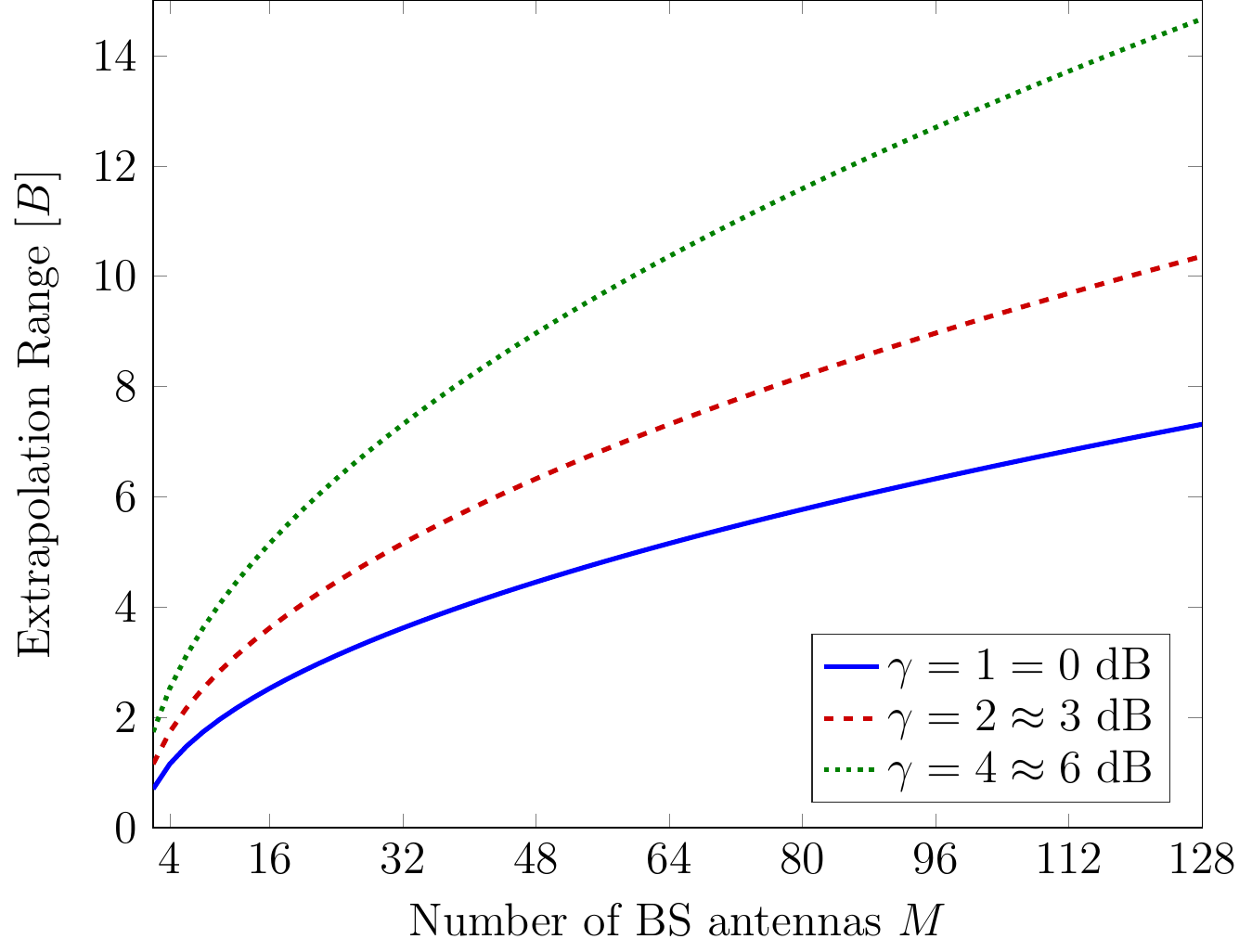}}
	}
	\caption{\small (a) As number of antennas increases, the CRLB reaches the CRLB formula for separated paths while achieving an additional SNR gain. (b) Extrapolation range $f_{\mathrm{Extrapol-\gamma}}$, quantified in multiples of training band $B$, is defined as the extrapolation range beyond which the CRLB falls below $\gamma$ times the one of the LS estimator in-band. It is based on the well separated CRLB formula.}
	\label{fig:impact_M} 

\end{figure*}

For the same set of channel parameters, Fig.~\ref{fig:impact_M} (a) depicts the extrapolation performance for different numbers of antenna elements. As the number of antennas increases, the resolution in the angle domain increases and the BS can better resolve the paths. This implies that $\mathbf{(As3)}$ becomes valid and the CRLB converges to the simplified CRLB. Note that, in the $M=4$ case, extrapolation performance in terms of MSE deteriorates quickly as we move away from the uplink band. The simplified CRLB of Proposition~\ref{proposition:well_separated_rays} is very close to the full CRLB as soon as the array has 16 antennas. Moreover, as the number of antennas $M$ increases, a corresponding in-band array gain is achieved implying that the curves are shifted 6 dB down as the number of antennas is multiplied by 4.

In Fig.~\ref{fig:impact_M} (b), the extrapolation range $f_{\mathrm{Extrapol-\gamma}}$, expressed in (\ref{eq:extrapolation_range}), is plotted as a function of the number of antennas. As a reminder, the formula corresponds to the extrapolation range $f$ beyond which the CRLB performance is $\gamma$ times worse than the one of the conventional LS estimator in-band. The formula is independent of the SNR and assumes that the paths are well separated, \textit{i.e.}, $\mathbf{(As3)}$ is valid. As an example, if the BS has 128 antennas and the uplink training band is $B=20$ MHz, the FDD separation between uplink and downlink bands should be at most $7B=140$ MHz to guarantee that the extrapolation-based FDD downlink MSE performance is equivalent to the corresponding one of a TDD system using conventional LS estimation.

\subsection{Spectral Efficiency}
\label{subsection:spectral_efficiency}

\begin{figure*}[t]
	\centering 
		\subfloat[ ]{
	\resizebox{0.49\textwidth}{!}{
	\includegraphics{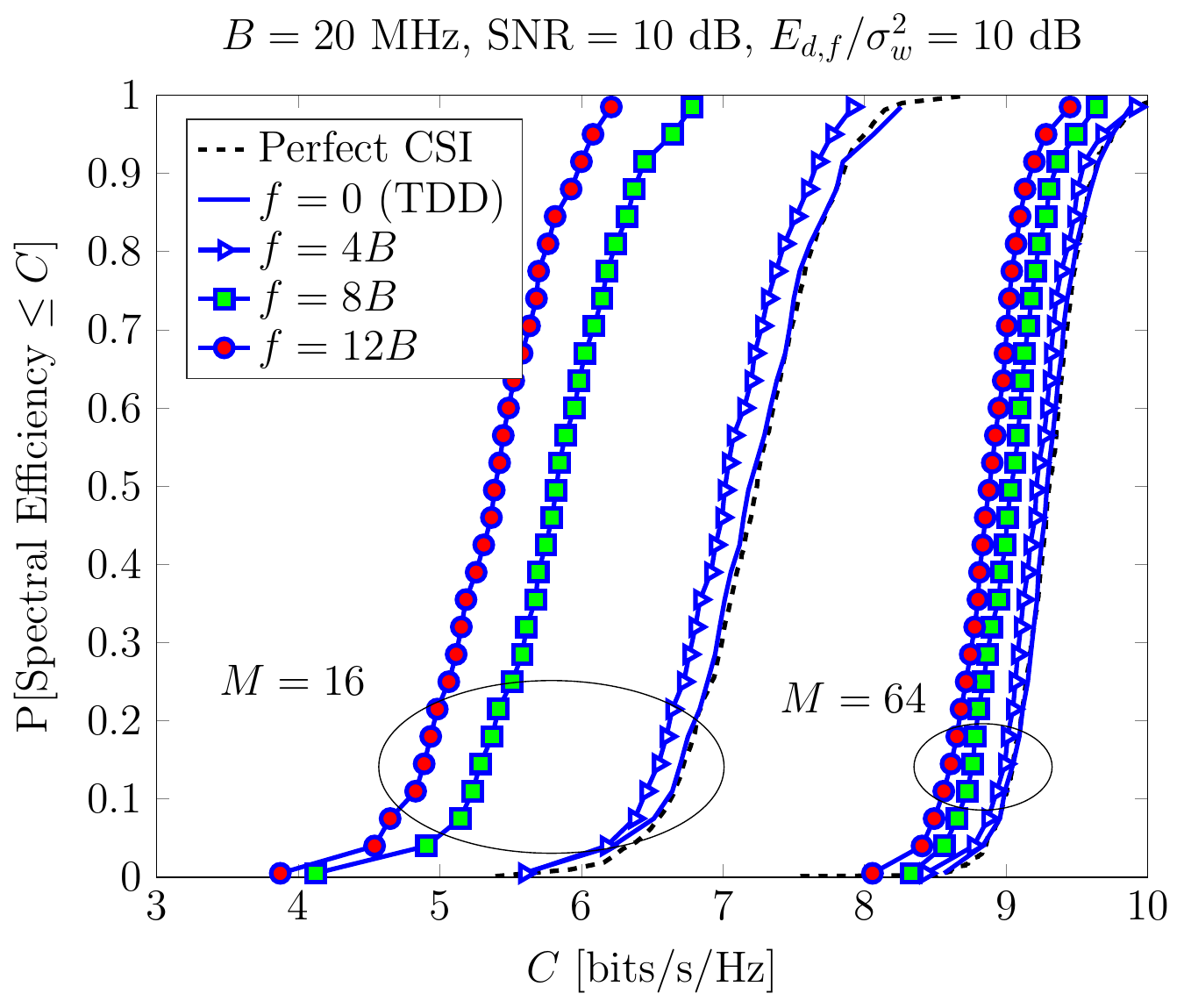}}
		}
		\subfloat[ ]{
			\resizebox{0.49\textwidth}{!}{
			\includegraphics{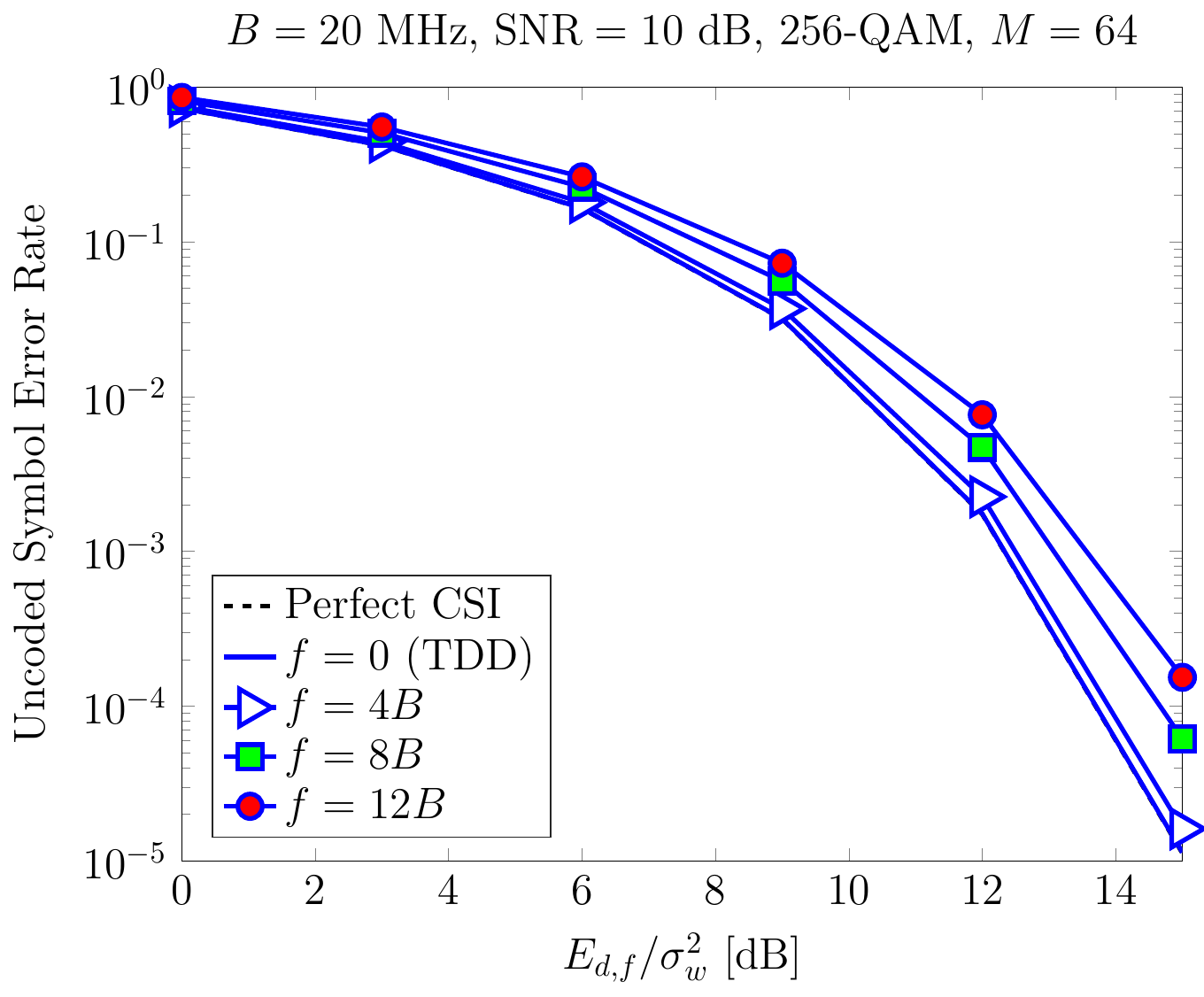}}
		}
	\caption{\small (a) Cumulative density function of the spectral efficiency of extrapolation-based FDD systems as compared to a TDD system, for different number of BS antennas $M$ and extrapolation frequency $f$ (separation between uplink carrier frequency and downlink subcarrier frequency). (b) Corresponding uncoded symbol error rate performance for $M=64$ antenna case.}
	\label{fig:CDF_spectral_efficiency} 
\end{figure*}

To evaluate the loss in spectral efficiency of an extrapolation-based FDD system versus a corresponding TDD system, we use the QuaDRIiGa toolbox to generate the channel parameters related to 200 users locations randomly distributed in a radius of 200 meters around the BS; the BS height is 20m above ground. The channel model used is still the 3GPP 3D-UMa NLOS model. For each user location, we compute the spectral efficiency at frequency $f$ according to (\ref{eq:spectral_efficiency}) where $\hat{\eta}(f)$ is computed based on (\ref{eq:efficiency_approx_simplified}) and $\mat{E}(f)=\mat{C}(f)$ given in (\ref{eq:CRLB_covariance_matrix}). The SNR related to uplink pilots is set to $\mathrm{SNR}=10$ dB and the downlink SNR on the received symbol is $E_{d,f}/\sigma_w^2=10$ dB as well. The uplink training bandwidth is left to 20 MHz.

Fig.~\ref{fig:CDF_spectral_efficiency} (a) plots the cumulative density function (CDF) of the spectral efficiency for two antenna settings $M=16$ and $M=64$ and for 4 values of the extrapolation frequency $f$: $f=0$ corresponds to the uplink carrier frequency, which is in-band and also corresponds to the downlink performance of a TDD system, $f\in \{80, 160, 240\}$ MHz corresponds to the FDD downlink performance of systems working at different extrapolation frequencies. The performance of a perfect CSI system is also plotted but can hardly be distinguished from the $f=0$ TDD performance. As the extrapolation frequency $f$ increases, the inaccurate CSI induces a beamforming power loss and a related loss of spectral efficiency for FDD systems as compared to TDD systems. Note that this loss is much more pronounced for the $M=16$ case than the $M=64$ case. This can be explained by the fact that the channel estimation and related extrapolation is more accurate in the $M=64$ case as improved spatial resolution is available. 

Fig.~\ref{fig:CDF_spectral_efficiency} (b) plots the uncoded symbol error rate performance, computed based on (\ref{eq:SER}) averaged over the different user locations, for the same parameters in the $M=64$ case. A 256-QAM constellation is considered. Here again, TDD performance ($f=0$) and perfect CSI curves fit. As the extrapolation frequency $f$ increases, the performance of the corresponding FDD system is degraded but still remains in a 2 dB range from the TDD performance.

\section{Conclusions}
\label{section:conclusion}

This paper investigated the performance of extrapolation-based FDD massive MIMO systems, relying on high-resolution parameter estimation. We demonstrated that, under a good calibration of the BS and favorable propagation conditions, channel extrapolation is a viable solution to deploy FDD massive MIMO systems. It has the great advantage to drastically reduce the DL pilot overhead and completely remove the need for a feedback from the users. The price to pay is a reduction in the quality of the channel estimates, which results in a performance loss in the user downlink transmission. Theoretical CRLB for the MSE of the extrapolated channel and the related user SNR performance were derived and validated through numerical simulations. Our simulation results show that extrapolation-based FDD systems relying on high-resolution channel estimation are a feasible and attractive solution, even as compared to a corresponding TDD system. In particular, we showed that the FDD performance only suffers from a 1 to 3 dB reduction in beamforming power for extrapolation range as large as 200 MHz for a BS equipped with 64 antennas.

Our future studies will include performance assessment of extensive outdoor measurements. In particular, the impact of calibration errors and channel modeling errors such as, \textit{e.g.}, dense multipath components, will require further investigation. Another interesting perspective is to take into account the impact of multiple antennas at the user side. Intuitively, this should help the extrapolation as we saw that the performance highly depends on the path separability in at least one domain, which helps their estimation as being free from the interference of other paths. In SIMO, paths can only be separated in the delay and angle of arrival domain while in MIMO they can be additionally separated in the angle of departure domain. Similarly, accounting for the time variations of the channel could be beneficial too as paths could be separated in the Doppler domain.

\section{Appendix}

Using (\ref{eq:Fisher_info_matrix_element}), we can compute the different elements of the Fisher information matrix given in (\ref{eq:Fisher_information_matrix}). In the following, we use the notations  $\|\vect{s}_l\|^2=\|\vect{s}\|^2$ and $\|\dot{\vect{s}}_l\|^2=\|\dot{\vect{s}}\|^2$ given that the dependence on the path index vanishes.

First, using $\mathbf{(As3)}$, we can show that the off-diagonal blocks of $\mat{I}_{\vect{\psi}}$ vanish, \textit{i.e.}, $\mat{I}_{\vect{\psi}_l,\vect{\psi}_{l'}}=\mat{0}$ for $l\neq l'$. Indeed, for the diagonal elements of $\mat{I}_{\vect{\psi}_l,\vect{\psi}_{l'}}$, we find that
\begin{align*}
I_{\tau_l \tau_{l'}}&=\Re \left( \alpha_l^*\alpha_{l'} \vect{a}_l^H\vect{a}_{l'}   
\dot{\vect{s}}^H_l\dot{\vect{s}}_{l'}\right)=|\alpha_l|^2 \|\vect{a}_l\|^2 \|\dot{\vect{s}}\|^2 \delta_{l-l'}\\
I_{\phi_l \phi_{l'}}&=\Re \left( \alpha_l^*\alpha_{l'} \dot{\vect{a}}_{l,\phi}^H \dot{\vect{a}}_{l',\phi} {\vect{s}}_{l}^H{\vect{s}}_{l'}\right)=|\alpha_l|^2 \|\dot{\vect{a}}_{l,\phi}\|^2 \|\vect{{s}}\|^2 \delta_{l-l'}\\
I_{\theta_l \theta_{l'}}&=\Re \left( \alpha_l^*\alpha_{l'} \dot{\vect{a}}_{l,\theta}^H \dot{\vect{a}}_{l',\theta} {\vect{s}}_{l}^H{\vect{s}}_{l'}\right)=|\alpha_l|^2 \|\dot{\vect{a}}_{l,\theta}\|^2 \|{\vect{s}}\|^2 \delta_{l-l'}\\
I_{\alpha^R_l \alpha^R_{l'}}&=I_{\alpha^I_l \alpha^I_{l'}}=\Re \left( \vect{a}_l^H \vect{a}_{l'} {\vect{s}}_{l}^H{\vect{s}}_{l'}\right)=\|\vect{a}_l\|^2 \|{\vect{s}}\|^2 \delta_{l-l'}.
\end{align*}
Still using $\mathbf{(As3)}$, we find the same results for the off-diagonal elements of $\mat{I}_{\vect{\psi}_l,\vect{\psi}_{l'}}, l\neq l'$. Actually, using $\mathbf{(As4)}$, we find that the result also holds when $l=l'$ for the following elements
\begin{align*}
I_{\tau_l \phi_{l'}}&=-\Re \left( \alpha_l^*\alpha_{l'} \vect{a}_l^H \dot{\vect{a}}_{l',\phi} \dot{\vect{s}}_{l}^H{\vect{s}}_{l'}\right)=0\\
I_{\tau_l \theta_{l'}}&=-\Re \left( \alpha_l^*\alpha_{l'} \vect{a}_l^H \dot{\vect{a}}_{l',\theta} \dot{\vect{s}}_{l}^H{\vect{s}}_{l'}\right)=0\\
I_{\tau_l \alpha^R_{l'}}&=-\Re \left( \alpha_l^* \vect{a}_l^H {\vect{a}}_{l',} \dot{\vect{s}}_{l}^H{\vect{s}}_{l'}\right)=0\\
I_{\tau_l \alpha^I_{l'}}&=\Im \left( \alpha_l^* \vect{a}_l^H {\vect{a}}_{l',} \dot{\vect{s}}_{l}^H{\vect{s}}_{l'}\right)=0\\
I_{\phi_l \alpha^R}&=\Re \left( \alpha_l^* \dot{\vect{a}}_{l,\phi}^H {\vect{a}}_{l'} {\vect{s}}_{l}^H{\vect{s}}_{l'}\right)=0\\
I_{\phi_l \alpha^I}&=-\Im \left( \alpha_l^* \dot{\vect{a}}_{l,\phi}^H {\vect{a}}_{l'} {\vect{s}}_{l}^H{\vect{s}}_{l'}\right)=0\\	
I_{\theta_l \alpha^R_{l'}}&=\Re \left( \alpha_l^* \dot{\vect{a}}_{l,\theta}^H {\vect{a}}_{l'} {\vect{s}}_{l}^H{\vect{s}}_{l'}\right)=0\\
I_{\theta_l \alpha^I_{l'}}&=-\Im \left( \alpha_l^* \dot{\vect{a}}_{l,\theta}^H {\vect{a}}_{l'} {\vect{s}}_{l}^H{\vect{s}}_{l'}\right)=0\\
I_{\alpha^R_l \alpha^I_{l'}}&=-\Im \left(\vect{a}_l^H \vect{a}_{l'} \vect{s}_{l}^H\vect{s}_{l'}\right)=0.
\end{align*}
One can further check that, under $\mathbf{(As3)}$, the elements $ I_{\phi_l \theta_{l'}}$ vanish for $l\neq l'$. However, even under $\mathbf{(As4)}$, $ I_{\phi_l \theta_{l'}}$ does not vanish for $l=l'$, \textit{i.e.},
\begin{align*}
I_{\phi_l \theta_{l'}}&=\Re \left( \alpha_l^*\alpha_{l'} \dot{\vect{a}}_{l,\phi}^H \dot{\vect{a}}_{l,\theta} {\vect{s}}_{l}^H{\vect{s}}_{l'}\right)= |\alpha_l|^2 \|{\vect{s}}\|^2 \Re\left( \dot{\vect{a}}_{l,\phi}^H \dot{\vect{a}}_{l,\theta}\right) \delta_{l-l'}.
\end{align*}
Taking into account the above simplifications, the full Fisher matrix $\mat{I}_{\vect{\psi}}$ becomes block diagonal and each block on its diagonal is itself block diagonal
\begin{align*}
\mat{I}_{\vect{\psi}}&=\frac{2}{\sigma_w^2}\begin{pmatrix}
\mat{I}_{\vect{\psi}_1,\vect{\psi}_1}&  & & \\
& \ddots & &\\
& & \ddots   &\\
& & & \mat{I}_{\vect{\psi}_L,\vect{\psi}_L}
\end{pmatrix},\\
%
\mat{I}_{\vect{\psi}_l,\vect{\psi}_{l}}&=\begin{pmatrix}
I_{\tau_l \tau_{l}}    &    &   &   & \\
& I_{\phi_l \phi_{l}} & I_{\phi_l \theta_{l}}  &  & \\
& I_{\phi_l \theta_{l}}   & I_{\theta \theta_{l}}      &    & \\
&  &   & I_{\alpha^R_l \alpha^R_{l}}   & \\
& &  & & I_{\alpha^I_l \alpha^I_{l}}\\
\end{pmatrix}. \nonumber
\end{align*}
Using the fact the inverse of a block diagonal matrix is a block diagonal matrix with the inverse of the original blocks on its diagonal, the CRLB of (\ref{eq:CRLB_MSE}) averaged over the receive antennas becomes
\begin{align}
&\mathrm{MSE}(f)\geq\frac{1}{M}\sum_{m=1}^{M}\vect{g}^H_{m,f,\vect{\psi}}\mat{I}_{\vect{\psi}}^{-1}\vect{g}_{m,f,\vect{\psi}}\nonumber\\
&=\frac{\sigma_w^2}{2M}\sum_{m=1}^{M}\sum_{l=1}^L\vect{g}^H_{m,f,\vect{\psi}_l}\mat{I}_{\vect{\psi}_l,\vect{\psi}_l}^{-1}\vect{g}_{m,f,\vect{\psi}_l}\label{eq:eq_LB}\\
&=\frac{\sigma_w^2}{2M}\sum_{m=1}^{M}\sum_{l=1}^L \left[ \frac{|g_{m,f,\tau_l}|^2}{I_{\tau_l \tau_l}}+\frac{|g_{m,f,\alpha^R_l}|^2}{I_{\alpha^R_l \alpha^R_l}}+\frac{|g_{m,f,\alpha^I_l}|^2}{I_{\alpha^I_l \alpha^I_l}}
\right.\nonumber\\
&\left. + \begin{pmatrix}
g_{m,f,\phi_l}^* & g_{m,f,\theta_l}^*
\end{pmatrix}\begin{pmatrix}
I_{\phi_l \phi_l}     & I_{\phi_l \theta_l}\nonumber\\
I_{\phi_l \theta_l}   & I_{\theta_l \theta_l}
\end{pmatrix}^{-1}\begin{pmatrix}
g_{m,f,\phi_l} \\ g_{m,f,\theta_l}
\end{pmatrix}\right] .\nonumber
\end{align}
After some computations, we find that
\begin{align*}
&\sum_{m=1}^{M}\sum_{l=1}^L \frac{|g_{m,f,\alpha^R_l}|^2}{I_{\alpha^R_l \alpha^R_l}}+\frac{|g_{m,f,\alpha^I_l}|^2}{I_{\alpha^I_l \alpha^I_l}}=\frac{2L}{\|\vect{s}\|^2} \\
&\sum_{m,l}  \begin{pmatrix}
g_{m,f,\phi_l}^* & g_{m,f,\theta_l}^*
\end{pmatrix}\begin{pmatrix}
I_{\phi_l \phi_l}     & I_{\phi_l \theta_l}\nonumber\\
I_{\phi_l \theta_l}   & I_{\theta_l \theta_l}
\end{pmatrix}^{-1}\begin{pmatrix}
g_{m,f,\phi_l} \\ g_{m,f,\theta_l}
\end{pmatrix}
=\frac{2L}{\|\vect{s}\|^2}\\
&\sum_{m=1}^{M}\sum_{l=1}^L \frac{|g_{m,f,\tau_l}|^2}{I_{\tau_l \tau_l}}=\frac{L(2\pi f)^2}{\|\dot{\vect{s}}\|^2} =\frac{L}{\|\vect{s}\|^2} \frac{f^2}{\sigma_F^2},
\end{align*}
where $\sigma_F^2=\frac{\|\dot{\vect{s}}\|^2}{(2\pi)^2\|\vect{s}\|^2}$. Inserting the result of these last equations into (\ref{eq:eq_LB}) and using the definition $E_T = \|\vect{s}\|^2$, we find the result of Proposition~\ref{proposition:well_separated_rays}.


\ifCLASSOPTIONcaptionsoff
  \newpage
\fi




\footnotesize
\bibliographystyle{IEEEtran}
\bibliography{IEEEabrv,refs}
\end{document}